\documentclass[aps,pre,twocolumn,superscriptaddress,longbibliography]{revtex4-1}

\usepackage{upgreek}
\usepackage{amsmath}
\usepackage{amssymb}
\usepackage{graphicx}
\usepackage{braket}
\usepackage{lipsum}
\usepackage{xcolor}
\usepackage{epsfig}
\usepackage{enumitem}
\usepackage{mathtools}
\usepackage[implicit=false]{hyperref}
\usepackage[outdir=./FiguresLabels/]{epstopdf}
\usepackage[caption=false]{subfig}
\graphicspath{{FiguresLabels/}}

\begin{document}
	
	\title{The nonlinear semiclassical dynamics of the unbalanced, open Dicke model}
	
	\author{Kevin C. Stitely}
	\email{ksti263@aucklanduni.ac.nz}
	\affiliation{Dodd-Walls Centre for Photonic and Quantum Technologies, New Zealand}
	\affiliation{Department of Physics, University of Auckland, Auckland 1010, New Zealand}
	\affiliation{Department of Mathematics, University of Auckland, Auckland 1010, New Zealand}
	\author{Andrus Giraldo}
	\author{Bernd Krauskopf}
	\affiliation{Dodd-Walls Centre for Photonic and Quantum Technologies, New Zealand}
	\affiliation{Department of Mathematics, University of Auckland, Auckland 1010, New Zealand}
	\author{Scott Parkins}
	\affiliation{Dodd-Walls Centre for Photonic and Quantum Technologies, New Zealand}
	\affiliation{Department of Physics, University of Auckland, Auckland 1010, New Zealand}
	
	\date{\today}
	
	
	\begin{abstract}

	In recent years there have been significant advances in the study of many-body interactions between atoms and light confined to optical cavities. One model which has received widespread attention of late is the Dicke model, which under certain conditions exhibits a quantum phase transition to a state in which the atoms collectively emit light into the cavity mode, known as superradiance. We consider a generalization of this model that features independently controllable strengths of the co- and counter-rotating terms of the interaction Hamiltonian. We study this system in the semiclassical (mean field) limit, i.e., neglecting the role of quantum fluctuations. Under this approximation, the model is described by a set of nonlinear differential equations, which determine the system's semiclassical evolution. By taking a dynamical systems approach, we perform a comprehensive analysis of these equations to reveal an abundance of novel and complex dynamics. Examples of the novel phenomena that we observe are the emergence of superradiant oscillations arising due to Hopf bifurcations, and the appearance of a pair of chaotic attractors arising from period-doubling cascades, followed by their collision to form a single, larger chaotic attractor via a sequence of infinitely many homoclinic bifurcations. Moreover, we find that a flip of the collective spin can result in the sudden emergence of chaotic dynamics. Overall, we provide a comprehensive roadmap of the possible dynamics that arise in the unbalanced, open Dicke model in the form of a phase diagram in the plane of the two interaction strengths. Hence, we lay out the foundations to make further advances in the study of the fingerprint of semiclassical chaos when considering the master equation of the unbalanced Dicke model, that is, the possibility of studying a manifestation of quantum chaos in a specific, experimentally realizable system.
		
	\end{abstract}
	
	
	\maketitle
	
	\section{Introduction}
	
	Since the advent of chaos theory in the 1960's, the question of how classical chaotic behavior arises from quantum dynamics has been a topic of significant interest. After all, the linearity of the formulation of quantum mechanics disallows the existence of exponential sensitivity to initial conditions that is characteristic of classical chaotic dynamics \cite{wimberger_nonlinear_2014}. In the late 20th century, the study of the onset of chaotic behavior in quantum mechanical models found progress in the field of many-body quantum electrodynamics (QED). More recently, quantum chaos has been extremely topical due to its connection with information scrambling \cite{alavirad_scrambling_2019,lewis-swan_unifying_2019} and black hole physics \cite{shenker_black_2014,sekino_fast_2008}; it has even been suggested to provide routes to experimentally testing the holographic principle of quantum gravity \cite{zohar_quantum_2015,swingle_measuring_2016}.
	
	A quantum mechanical model that has recently brought attention to a plethora of uniquely many-body effects is the \emph{Dicke model}. This paradigmatic model of quantum optics describes the interaction of an ensemble of $N$ two-level atoms with a light field confined to an optical cavity \cite{dicke_coherence_1954,garraway_dicke_2011,kirton_introduction_2019}. The Dicke model has been of particular interest because of a quantum phase transition that the model can exhibit due to the onset of significant cooperative interactions of the atoms with the light field \cite{kirton_introduction_2019,grimsmo_dissipative_2013,bhaseen_dynamics_2012}. If the strength of the atom-field coupling is sufficiently high, the system can undergo a quantum phase transition from the so-called normal phase into a superradiant phase, which features a non-zero photon number at equilibrium \cite{hepp_equilibrium_1973,hepp_superradiant_1973,wang_phase_1973,klinder_dynamical_2015}.
	
	The Dicke model has been used to study chaos in many-body QED in the semiclassical regime \cite{hopf_chaos_1983,muller_classical_1991,finney_quantum_1996,finney_quasiclassical_1994} and in relation to the superradiant phase transition in a seminal pair of papers by Emary and Brandes \cite{emary_chaos_2003,emary_quantum_2003}. However, these treatments depend critically upon the Hamiltonian formulation of quantum mechanics and are, thus, limited to closed quantum systems. A more realistic approach that includes dissipative effects introduces a quantum master equation \cite{kirton_introduction_2019}; such a model is referred to as the open Dicke model. Recently, chaotic behavior was uncovered in a driven version of this model \cite{kirton_superradiant_2018}.
	
	The superradiant quantum phase transition described by the Dicke model has proved difficult to realize experimentally. However, technological advancements over the past decade have led to a number of successful realizations. Baumann \emph{et al.} \cite{baumann_dicke_2010,baumann_exploring_2011} and Klinder \emph{et al.} \cite{klinder_dynamical_2015} have produced an experimental realization of the Dicke model with a Bose--Einstein condensate coupled to an optical cavity. With this setup, the Dicke superradiant phase transition is produced when the Bose--Einstein condensate undergoes a symmetry breaking self-organization into a supersolid phase. Other realizations have been achieved by Hamner \emph{et al.} with a spin-orbit-coupled Bose--Einstein condensate confined in a harmonic trap \cite{hamner_dicke-type_2014} and by Safavi-Naini \emph{et al.} with a trapped-ion array \cite{safavi-naini_verification_2018}.
	
	\begin{figure}[h]
		\centering
		\includegraphics[width=8.6cm]{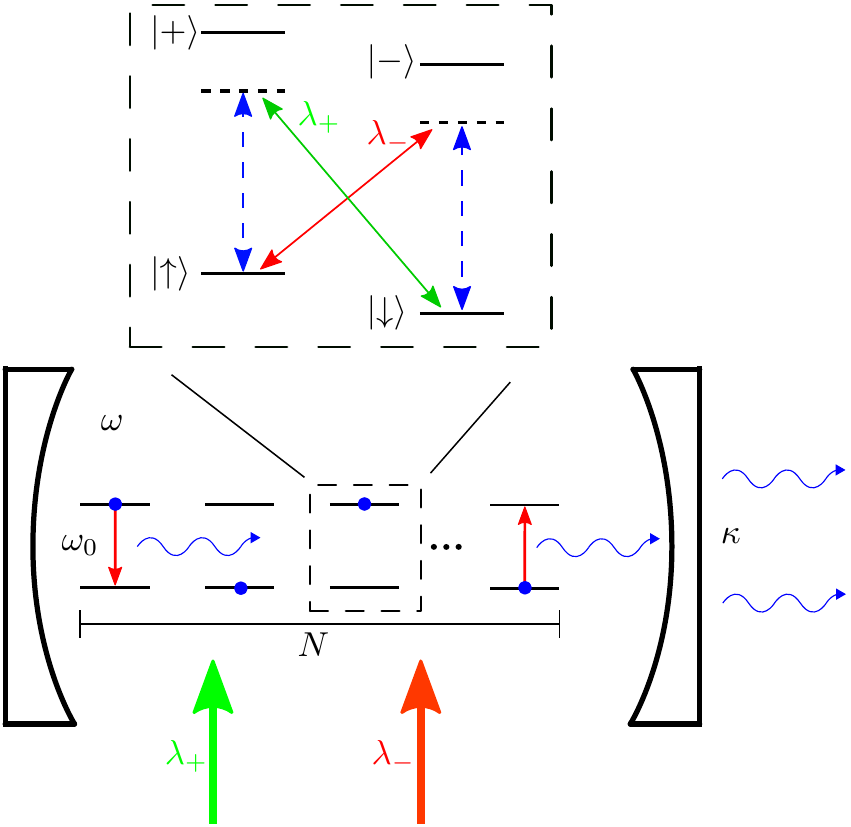}
		\caption{\label{fig:Schematic}Schematic of the unbalanced, open Dicke model as proposed by Dimer \emph{et al.} \cite{dimer_proposed_2007} in a cavity QED system. An optical cavity with a resonant frequency $\omega$ and decay rate $\kappa$ is coupled to $N$ atoms with frequency splitting $\omega_0$. The atoms are also driven by two independent lasers, which produce two independent Raman coupling strengths $\lambda_{\pm}$. Each atom is a four-level system with two stable ground states $\ket{\downarrow}$ and $\ket{\uparrow}$. The laser and cavity fields are sufficiently detuned from the atomic transitions that the two excited states $\ket{+}$ and $\ket{-}$ can be adiabatically eliminated in the theoretical model, producing an effective two-level system with comparable rotating and counter-rotating atom-cavity interaction terms.}
	\end{figure}
	
	On the optical side, Dimer \emph{et al.} \cite{dimer_proposed_2007} proposed engineering an effective Dicke model in a cavity QED experiment with atom-photon interactions mediated by Raman transitions between ground atomic spin states of a four-level system. This scheme is illustrated in Fig. \ref{fig:Schematic}. The proposal considers a four-level atom with two ground spin states, $\ket{\downarrow}$ and $\ket{\uparrow}$, and two excited states $\ket{-}$ and $\ket{+}$, highly detuned from the cavity and laser frequencies. Two independent lasers drive the transitions $\ket{\uparrow}\leftrightarrow\ket{-}$ and  $\ket{\downarrow}\leftrightarrow\ket{+}$, respectively, while the cavity mode couples simultaneously to the transitions $\ket{\uparrow}\leftrightarrow\ket{+}$ and $\ket{\downarrow}\leftrightarrow\ket{-}$. Together, the lasers and cavity mode drive two, independent Raman transitions between the ground spin states and, if the detunings of the fields from the excited states are sufficiently large, these states can be adiabatically eliminated to yield an effective two-state-atom model. Significantly, in this effective model counter-rotating and co-rotating interactions feature on an equal footing; i.e., the excitation of the atom from $\ket{\downarrow}$ to $\ket{\uparrow}$ may occur with either absorption of a photon from, or emission of a photon into, the cavity mode, as a result of the two possible Raman transition paths between $\ket{\downarrow}$ and $\ket{\uparrow}$.
	
	The generalization of the Dicke model to feature two independent coupling strengths for the co- and counter-rotating interactions, which we refer to as the \emph{unbalanced, open Dicke model}, is the focus of this paper. It was recently realized by Zhiqiang \emph{et al.}, whose experimental and theoretical results have shown that this generalization yields interesting nonequilibrium phase behavior and, in particular, oscillatory phases \cite{zhiqiang_nonequilibrium_2017,soriente_dissipation-induced_2018}.
	
	In the work presented here, we take a dynamical systems approach to analyze the set of nonlinear differential equations that arise from the semiclassical model. In doing so, we demonstrate that, in contrast to equal coupling, the option of unbalanced coupling reveals a wide variety of dynamical behavior. In particular, we analyze in detail the semiclassical description of a superradiant phase transition that occurs in two stages, corresponding to the onset of superradiance and the subsequent disappearance of the normal phase. In between the two stages, the system features a phase coexistence of the normal and superradiant phases, as noted in \cite{soriente_dissipation-induced_2018}. We are also able to use our approach to explain the results of Zhiqiang \emph{et al.} \cite{zhiqiang_nonequilibrium_2017}, showing the appearance of superradiant oscillations as arising from Hopf bifurcations. In addition to this, we describe a new type of transition that occurs between the two types of normal phases; spin-down, where the atomic population is entirely de-excited, or spin-up, where it is entirely excited. We then show that, at the transition point, an infinite set of energy-conserving periodic orbits emerge on the Bloch sphere for non-zero dissipation rates. We also analyze the onset of chaos in the system, demonstrating period-doubling cascades and collisions of chaotic attractors. To summarize all of the behaviors described above, we present an overview of the possible dynamics of the model in the form of a phase diagram in the plane of the two coupling strengths that appear due to the generalization of unbalanced coupling. Our exposition provides an original and comprehensive, nonlinear dynamics perspective on the unbalanced Dicke model, complementing other recent interest in this quite fascinating system \cite{soriente_dissipation-induced_2018,shchadilova_fermionic_2020}. 
	
	The paper is organized as follows. In Sect. \ref{Sect:UnbalancedDickeModel}, we outline the unbalanced, open Dicke model and its semiclassical approximation as a set of ordinary differential equations. In Sect. \ref{Sect:PhysicalAttractors}, we describe the possible dynamical phases and their mathematical description as solutions to said differential equations in the semiclassical regime. In Sect. \ref{Sect:SteadyStateBifurcations}, we present a bifurcation analysis of the steady-states and their multistability, where we demonstrate the two-stage superradiant phase transition. In Sect. \ref{Sect:Multistability}, we discuss multistability between phases in more detail and show the basins of attraction for the normal and superradiant phases on the Bloch sphere for an initially empty cavity. In Sect. \ref{Sect:Oscillations}, we analyze the emergence of superradiant oscillations, as demonstrated experimentally in \cite{zhiqiang_nonequilibrium_2017}. In Sect. \ref{Sect:PoleFlipTransition}, we explore the transition that arises between the two types of normal phases, before moving onto an analysis of the emergence of chaotic dynamics in Sect. \ref{Sect:Chaos}. Conclusions and an outlook are presented in Sect. \ref{Sect:Conclusion}.

	\section{Unbalanced Dicke model}\label{Sect:UnbalancedDickeModel}
	
	We consider an ensemble of $N$ identical two-level atoms interacting cooperatively with a single mode of the radiation field. The unbalanced Dicke model features two different coupling strengths, $\lambda_{\pm}$, for the co- and counter-rotating terms in the Hamiltonian (with $\hbar = 1$),
	\begin{align}\label{eqn:Hamiltonian}
	\hat{H} =&\  \omega \hat{a}^{\dagger}\hat{a} + \omega_0 \hat{J}_{z}  +  \frac{\lambda_{-}}{\sqrt{N}}\left( \hat{a}\hat{J}_{+} + \hat{a}^{\dagger}\hat{J}_{-} \right) \nonumber \\
	&+ \frac{\lambda_{+}}{\sqrt{N}}\left( \hat{a}\hat{J}_{-} + \hat{a}^{\dagger}\hat{J}_{+} \right),
	\end{align}
	where $\hat{a}$ is the annihilation operator of the cavity field mode, $\omega$ is the frequency of the field mode, and $\omega_0$ is the frequency splitting between the atomic levels.
	
	$\hat{J}_{\pm,z}$ are the collective angular momentum operators, given by
	\begin{equation}
	\hat{J}_{\pm,z} = \sum_{\nu=1}^{N}\hat{\sigma}_{\pm,z}^{(\nu)}\ ,
	\end{equation}
	where $\hat{\sigma}_{\pm,z}^{(\nu)}$ are the (spin-1/2) angular momentum operators for the $\nu$th atom. The collective operators form an angular momentum algebra such that $[\hat{J}_{+},\hat{J}_{-}] = 2\hat{J}_{z}$, $[\hat{J}_{z},\hat{J}_{\pm}] = \pm \hat{J}_{\pm}$, and they have a linear degenerate subspace spanning the $2^N$-dimensional Hilbert space \cite{garraway_dicke_2011}. A basis is $\{\ket{J,M}\}$, where
	\begin{align}
	(\hat{J}_{x}^2 + \hat{J}_{y}^2 + \hat{J}_{z}^2)\ket{J,M} &= J(J+1)\ket{J,M}, \\
	\hat{J}_{z}\ket{J,M} &= M\ket{J,M},
	\end{align}
	with
	\begin{align}
	J &\in \{N/2,N/2-1,\cdots,-N/2+1,-N/2\}, \\
	M &\in \{J,J-1,\cdots,-J+1,-J \}.
	\end{align}
	
	The states $\ket{J,M}$ are referred to as the \emph{Dicke states}. Here we consider the case of maximal angular momentum, $J=N/2$. This has the effect of limiting the dimensionality of the collective atomic state to $N+1$ as only states with the same angular momentum are coupled. In particular, the $J=N/2$ subspace is referred to as the \emph{Dicke manifold} \cite{kirton_introduction_2019}. The Hilbert space of the entire system is then spanned by the basis $\{\ket{n}\otimes\ket{J,M}\}$, where $\ket{n}$ is a Fock state of the cavity mode defined by $\hat a^{\dagger}\hat a\ket{n}=n\ket{n}$.
	
	In the experiments of \cite{zhiqiang_nonequilibrium_2017}, the dominant source of dissipation is cavity loss. We model this with the standard quantum optical master equation in Lindblad form \cite{dimer_proposed_2007,zhiqiang_nonequilibrium_2017}
	\begin{equation}\label{eqn:MasterEqn}
	\frac{d\hat \rho}{dt} = -i[\hat H,\hat \rho] + \kappa\left( 2\hat a\hat \rho \hat a^{\dagger} - \hat a^{\dagger}\hat a\hat \rho - \hat \rho \hat a^{\dagger}\hat a \right),
	\end{equation}
	where $\kappa$ is the cavity field decay rate and $\hat \rho$ is the reduced density operator of the atom-cavity system.
	
	The Hamiltonian (\ref{eqn:Hamiltonian}) reduces to the usual, balanced Dicke Hamiltonian when $\lambda_{-}=\lambda_{+}$ and reduces to the Tavis-Cummings Hamiltonian when $\lambda_{+}=0$.
	
	\subsection{Semiclassical model}
	In the thermodynamic limit of $N\rightarrow\infty$, we consider the time-evolution of the expectation values of the operators $\hat a$, $\hat J_{-}$, and $\hat J_{z}$, scaled as
	\begin{equation}
	\alpha = \frac{\braket{\hat a}}{\sqrt{N}}\in \mathbb{C},\ \beta = \frac{\braket{\hat J_{-}}}{N} \in \mathbb{C},\ \gamma = \frac{\braket{\hat J_{z}}}{N}\in\mathbb{R},
	\end{equation}
	where $\gamma\in\mathbb{R}$ because $\hat J_{z}$ is a Hermitian operator. We use (\ref{eqn:Hamiltonian}) and (\ref{eqn:MasterEqn}) to derive a system of nonlinear differential equations for $\alpha$, $\beta$, and $\gamma$, given by
	\begin{subequations}\label{eqn:SemiclassicalEqns}
		\begin{align}
		\frac{d\alpha}{dt} &= -\kappa\alpha - i\omega\alpha - i\lambda_{-}\beta - i\lambda_{+}\beta^{*},\\
		\frac{d\beta}{dt} &= -i\omega_0\beta + 2i\lambda_{-}\alpha\gamma + 2i\lambda_{+}\alpha^{*}\gamma\ , \\
		\frac{d\gamma}{dt} &= i\lambda_{-}\left( \alpha^{*}\beta - \alpha\beta^{*} \right) + i\lambda_{+}\left( \alpha\beta - \alpha^{*}\beta^{*} \right).
		\end{align}
	\end{subequations}
	These equations are the semiclassical approximation of the unbalanced, open Dicke model. They form a five-dimensional dynamical system whose behavior we wish to explore.
	
	In the case of maximal angular momentum, the model exhibits spin conservation of the form
	\begin{equation}\label{eqn:SpinConservation}
	|\beta|^2 + \gamma^2 = \frac{1}{4},
	\end{equation}
	which confines the $\beta$- and $\gamma$-components of trajectories onto the Bloch sphere.
	
	Hence, the conservation law (\ref{eqn:SpinConservation}) eliminates a degree of freedom. Therefore, the system is four-dimensional with state space $\mathbb{R}^2\times\mathbb{S}^2$, which is a four-dimensional hypercylinder.
	
	\subsection{Symmetry of the unbalanced, open Dicke model}
	
	In addition to spin conservation, the unbalanced model also features a parity symmetry with the parity operator
	\begin{equation}
	\hat{\Pi} = e^{i\pi(\hat{a}^{\dagger}\hat{a} + \hat{J}_{z} + J)},
	\end{equation}
	where $\hat{\Pi}^{\dagger}\hat{a}\hat{\Pi}=-\hat{a}$, $\hat{\Pi}^{\dagger}\hat{J}_{-}\hat{\Pi} = -\hat{J}_{-}$, and $\hat{\Pi}^{\dagger}\hat{J}_{z}\hat{\Pi} = \hat{J}_{z}$. The unbalanced Dicke Hamiltonian (\ref{eqn:Hamiltonian}) is invariant under this parity transformation, i.e., $[\hat{H},\hat{\Pi}] = 0$. Furthermore, the Lindblad master equation (\ref{eqn:MasterEqn}) is invariant under the parity transformation. Hence, the full open system is $\mathbb{Z}_{2}$-invariant \cite{albert_symmetries_2014}.
	
	In the semiclassical regime, the parity symmetry manifests itself as a $\mathbb{Z}_{2}$-equivariance of system (\ref{eqn:SemiclassicalEqns}), given by the linear transformation
	\begin{equation}\label{eqn:Symmetry}
	T_{\mathbb{Z}_{2}}: \left( \alpha,\beta,\gamma \right)\rightarrow \left( -\alpha,-\beta,\gamma \right).
	\end{equation}
	
	In particular, this symmetry produces trajectories with a symmetric counterpart on the Bloch sphere. However, symmetry-related trajectories represent the same physically observable states, since the observables $|\alpha|^2$, $|\beta|^2$, and $\gamma$ are invariant under $T_{\mathbb{Z}_{2}}$.

	In addition to the parity symmetry (\ref{eqn:Symmetry}), the unbalanced, open Dicke model also features a parameter exchange symmetry, such that the exchange 
	\begin{equation}\label{eqn:ParameterSymmetry}
	T_{p}:(\omega_0,\lambda_{-},\lambda_{+})\rightarrow(-\omega_0,\lambda_{+},\lambda_{-}),
	\end{equation} 
	leaves the dynamics invariant subject to a reversal of the roles of the North and South poles of the Bloch sphere. Hence, the dynamics that occur near the South pole take place on the North pole under the parameter exchange (\ref{eqn:ParameterSymmetry}). The physical interpretation of the parameter exchange symmetry is that the rotating frame about which the co- and counter-rotating terms oscillate has changed direction. Thus, the ``co-rotating terms" are now rotating against the frame and vice versa for the counter-rotating terms. Then for the case $\omega_0<0$, $\omega>0$ the meaning of the co- and counter-rotating terms is reversed. Conversely, the associated exchange of $\lambda_{-}$ and $\lambda_{+}$ also implies a change in sign of the effective atomic frequency $\omega_0$.
	
	\subsection{Continuation and Stereographic Transformation}
	
	All computations are performed with the numerical continuation software package \textsc{auto-07p} \cite{doedel_auto:_1981,doedel_auto-07p:_2010}, which is used to find and continue equilibria, periodic orbits, and their bifurcations as system parameters are varied.
	
	However, the spherical part of the state space $\mathbb{R}^2\times\mathbb{S}^2$ of system (\ref{eqn:SemiclassicalEqns}) can introduce issues when attempting to continue solutions in the presence of conserved quantities. To remedy this, we transform the Bloch sphere to a plane by stereographic projection, such that the conservation law (\ref{eqn:SpinConservation}) is always implicitly satisfied.

	More specifically, we map every point $(\beta,\gamma)$ on the Bloch sphere $\mathbb{S}^2$, except for the chosen projection point, to the plane; hence, the transformed variables can take on any value in $\mathbb{R}^2$.
	
	With the real-valued variables
	\begin{align}
	a_{1} &= \mathrm{Re}(\alpha),\ a_{2} = \mathrm{Im}(\alpha), \\
	b_{1} &= \mathrm{Re}(\beta),\ b_{2} = \mathrm{Im}(\beta),
	\end{align}
	we choose the ``North pole,"
	\begin{equation}
	p=\{(b_{1},b_{2},\gamma)\in\mathbb{S}^2:b_{1}=b_{2}=0,\gamma=1/2\},
	\end{equation}
	of the Bloch sphere to be the projection point, and define the stereographic map
	\begin{align}
	g:\mathbb{S}^2\backslash p &\mapsto\mathbb{R}^2, \\
	(b_{1},b_{2},\gamma)&\mapsto (x,y)\equiv \left( \frac{-b_{1}}{\gamma-1/2}, \frac{-b_{2}}{\gamma - 1/2} \right).
	\end{align}
	Its inverse is
	\begin{equation}
	g^{-1}(x,y) = \left( \frac{x}{x^2 + y^2 +1},\frac{y}{x^2 + y^2 + 1},\frac{x^2 + y^2 - 1}{2x^2 + 2y^2 + 2} \right).
	\end{equation}
	
	The equations of motion for the transformed system in the variables $\{a_{1},a_{2},x,y\}$ are obtained by multiplication of the Jacobian matrix of the coordinate change $g$ by the vector containing the equations of motion for $\{a_{1},a_{2},b_{1},b_{2},\gamma\}$, then substituting the expressions for $\{ b_{1},b_{2},\gamma \}$ by $\{x,y\}$ (for more information see \cite{giraldo_saddle_2017}). The resulting equations of motion for the stereographically projected system on $\mathbb{R}^{4}$ are
	\begin{subequations}\label{eqn:StereoEqns}
		\begin{align}
		\begin{split}
		\frac{da_{1}}{dt} ={}& -\kappa a_{1}+\omega a_{2} + \frac{(\lambda_{-}-\lambda_{+})y}{x^2 + y^2 + 1},
		\end{split}\\
		\begin{split}
		\frac{da_{2}}{dt} ={}& -\omega a_{1} -\kappa a_{2}-\frac{(\lambda_{-}+\lambda_{+})x}{x^2+y^2+1},
		\end{split}\\
		\begin{split}
		\frac{dx}{dt}\hspace{1mm} ={}& (\lambda_{-} - \lambda_{+})a_{2}x^2 - 2(\lambda_{-}+\lambda_{+})a_{1}xy\\
		&+ (\lambda_{+}-\lambda_{-})a_{2}y^2+(\lambda_{-}-\lambda_{+})a_{2}+\omega_0 y,
		\end{split}\\
		\begin{split}
		\frac{dy}{dt}\hspace{1mm} ={}& (\lambda_{+}+\lambda_{-})a_{1}x^2+2(\lambda_{-}-\lambda_{+})a_{2}xy \\
		&- (\lambda_{+}+\lambda_{-})a_{1}y^2 - (\lambda_{-}+\lambda_{+})a_{1} - \omega_0 x.
		\end{split}
		\end{align}
	\end{subequations}
	
	\begin{figure}[h]
		\centering
		\includegraphics[width=8.6cm]{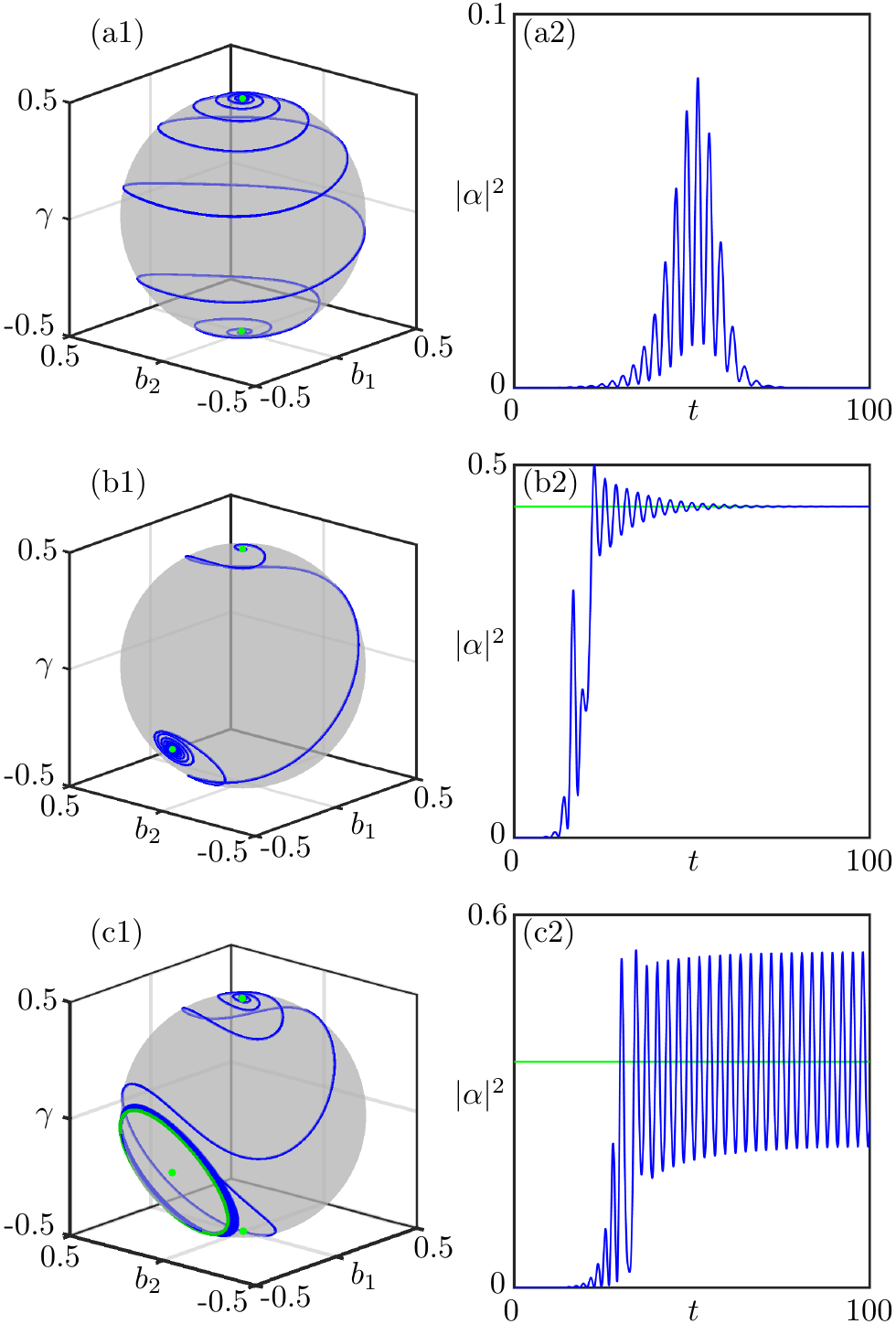}
		\caption{\label{fig:phases}Trajectories on the Bloch sphere and temporal traces of the photon number $|\alpha|^2$ showing the normal phase for $\lambda_{-} = \lambda_{+} = 0.4$ (a1-a2), superradiant phase for $\lambda_{-} = \lambda_{+} = 1$ (b1-b2), and oscillatory phase for $\lambda_{-} = 1$, $\lambda_{+} = 1.45$ (c1-c2). Other parameters are $\kappa = \omega = \omega_0 = 1$. The initial condition is $(\alpha,\beta,\gamma)=(0.001,0.001,0.4999)$.}
	\end{figure}

	System (\ref{eqn:StereoEqns}) is the central object of study of this paper; all bifurcation analysis is performed with these equations. Images of the Bloch sphere are obtained with the inverse transformation $g^{-1}$. Note that the $\mathbb{Z}_{2}$-equivariance now manifests itself as the symmetry transformation
	\begin{equation}\label{eqn:StereoParityTrans}
	T_{\mathbb{Z}_{2}}:(a_{1},a_{2},x,y)\rightarrow (-a_{1},-a_{2},-x,-y),
	\end{equation}
	which is the point reflection in the origin of $\mathbb{R}^4$.
	
	
	\section{Physical attractors}\label{Sect:PhysicalAttractors}
	
	A key result of the experiments performed by Zhiqiang \emph{et al.} \cite{zhiqiang_nonequilibrium_2017} is the existence of three phases of behavior: normal, superradiant, and oscillatory.
	The normal phase features a zero photon number in the steady-state, $\langle \hat{a}^{\dagger}\hat{a} \rangle\rightarrow 0$ as $t\rightarrow\infty$, and trivial atomic states, either $\ket{N/2,N/2}$ for spin-up or $\ket{N/2,-N/2}$ for spin-down. That is, all atoms are either excited or in their ground states. The superradiant phase, on the other hand, features two asymmetric steady-states, with non-zero photon number. The oscillatory phase is identified by persistent oscillations in the photon number about an unstable superradiant state.

	In the semiclassical regime, the normal phase is characterized by a stable equilibrium point with zero photon number, whereas the superradiant phase corresponds to a pair of stable equilibria with non-zero photon number. The oscillatory phase is marked by the existence of a pair of stable periodic orbits oscillating about the two unstable superradiant equilibria. Thus, we classify equilibria of the system as normal phase equilibria or superradiant equilibria. Figure \ref{fig:phases} shows a selection of characteristic trajectories on the Bloch sphere (left column) and temporal traces of the photon number (right column) for each of the three cases. In Figure \ref{fig:phases}(a1), we show a single trajectory from the North pole to the South pole (i.e., $\gamma\rightarrow -1/2$ as $t\rightarrow\infty$) in the normal phase, during which the cavity emits a pulse of photons, shown in Fig. \ref{fig:phases}(a2). In the superradiant phase, the symmetry of the steady-state is broken and the steady-state on the Bloch sphere is located off of the poles as shown in panel (b1). Here the photon number, shown in panel (b2), tends to a non-zero value. In the oscillatory phase, a pair of periodic orbits develop on the Bloch sphere, one of which is shown in panel (c1), and the photon number oscillates about a superradiant state, shown in panel (c2).

	
	\section{Steady-state bifurcations}\label{Sect:SteadyStateBifurcations}
	We first consider the effect of the generalization to unbalanced coupling on the superradiant phase transition. Here, we let the coupling strengths $\lambda_{\pm}$ vary while keeping $\kappa$, $\omega$, and $\omega_0$ fixed. This phase transition marks the onset of significant cooperative interactions between the ensemble of atoms and the radiation field. In this section, we discuss the superradiant phase transition in terms of the steady-state bifurcations of the semiclassical model (\ref{eqn:StereoEqns}).
	
\begin{figure}[t]
		\centering
		\includegraphics[width=8.6cm]{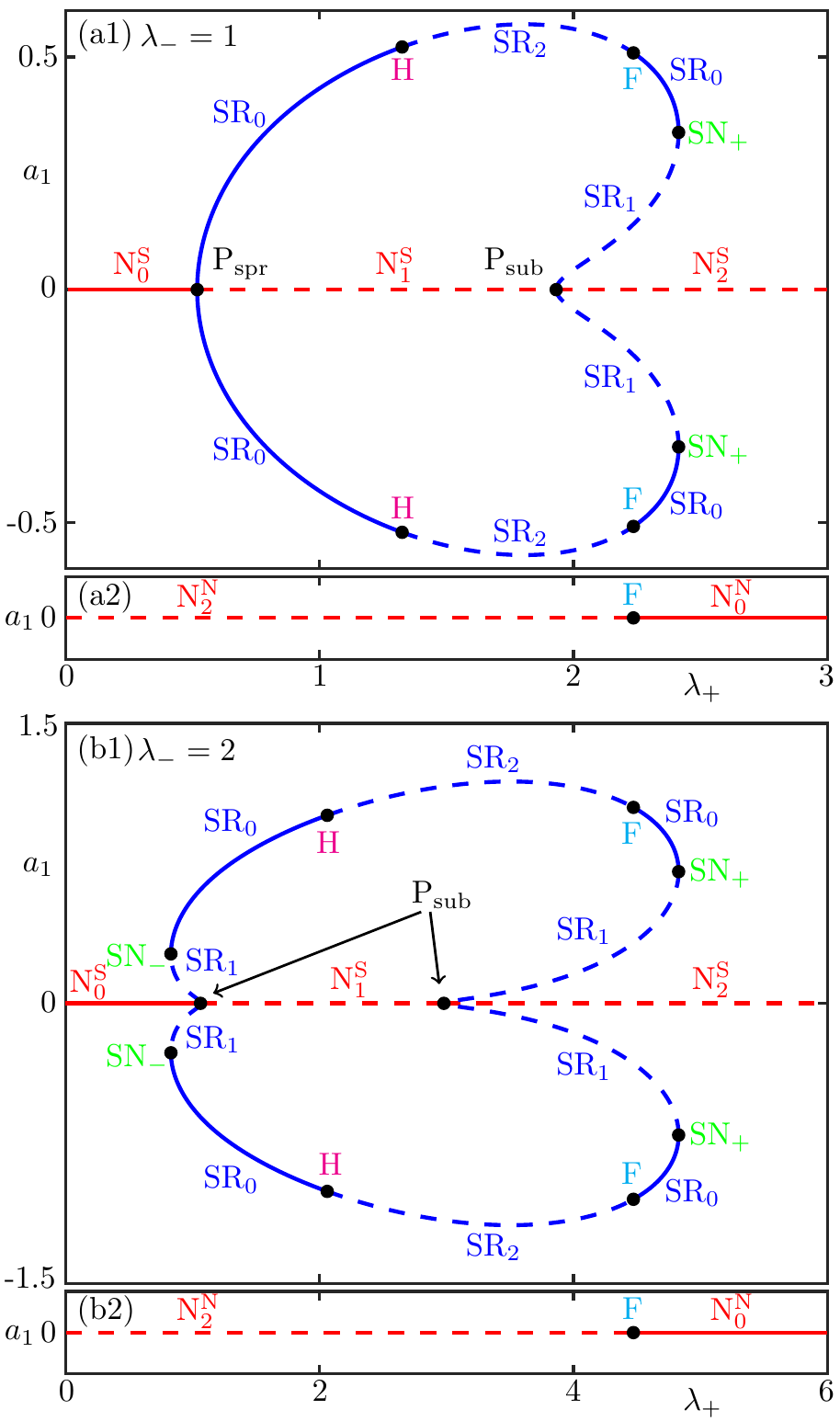}
		\caption{\label{fig:SprRadBifDiag}Bifurcation diagrams for the two scenarios of the superradiant phase transition. Panel (a) shows the one-stage onset of superradiance, and (b) shows the two-stage case. Solid curves indicate stable equilibria, dashed curves unstable equilibria and saddles. For (a1-a2) $\lambda_{-}=1$, for (b1-b2) $\lambda_{-}=2$; other parameters are $\kappa=\omega=\omega_0 = 1$. Normal phase equilibria are labelled N, and superscripts indicate whether the equilibrium is the North or South pole with N or S. Superradiant equilibria are labelled SR. Subscripts indicate the number of unstable directions.}
	\end{figure}

	We show in Fig. \ref{fig:SprRadBifDiag} the equilibria of system (\ref{eqn:StereoEqns}) in a bifurcation diagram for $\lambda_{-}=1$ in panels (a) and $\lambda_{-}=2$ in panels (b).
	Stable equilibria are shown as solid curves, whilst unstable equilibria are shown dashed. Normal phase equilibria are labelled N. The normal phase always has two equilibria which are distinguished with a superscript N for the North pole or S for the South pole. Superradiant equilibria are labelled SR. Subscripts indicate the number of unstable directions.
	
	In Fig. \ref{fig:SprRadBifDiag}(a), we show the bifurcation diagram for $\lambda_{-} = 1$. The bifurcations of the South pole and superradiant equilibria are given in Fig. \ref{fig:SprRadBifDiag}(a1), and the bifurcations of the North pole are shown in Fig. \ref{fig:SprRadBifDiag}(a2). The system starts, for $\lambda_{+}$ small, on the stable branch $\mathrm{N}_{0}^{\mathrm{S}}$ which then undergoes a supercritical pitchfork bifurcation $\mathrm{P}_{\mathrm{spr}}$ to become unstable along the branch $\mathrm{N}_{1}^{\mathrm{S}}$. Here, two branches of stable superradiant equilibria $\mathrm{SR}_0$ emerge. Each of them then becomes unstable at (a pair of) Hopf bifurcations H, and regains stability at the same time as the North pole equilibrium point at $\lambda_{+}\approx 2.236$; see Fig. \ref{fig:SprRadBifDiag}(a2). This transition marks a special point, labelled F, where the North pole becomes stable. This transition is discussed in more detail in Sect. \ref{Sect:PoleFlipTransition}. Slightly before this transition, at $\lambda_{+}\approx 1.932$, the unstable branch $\mathrm{N}_{1}^{\mathrm{S}}$ undergoes a subcritical pitchfork bifurcation $\mathrm{P}_{\mathrm{sub}}$ leading to the creation of two unstable branches $\mathrm{SR}_1$, which then meet the stable branches $\mathrm{SR}_0$ and vanish in the saddle-node bifurcations $\mathrm{SN}_+$.
	
	When $\lambda_{-} = 2$, shown in Fig. \ref{fig:SprRadBifDiag}(b), the transition to superradiance emerges in two stages. Again starting for $\lambda_{+}$ small, the system starts along the (stable) branch $\mathrm{N}_{0}^{\mathrm{S}}$. As $\lambda_{+}$ increases, a pair of saddle-node bifurcations $\mathrm{SN}_{-}$ create two superradiant branches each, the unstable branch $\mathrm{SR}_{\mathrm{1}}$ and the stable branch $\mathrm{SR}_{\mathrm{0}}$. The unstable branch $\mathrm{SR}_{\mathrm{1}}$ moves down to the South pole until it collides with the branch $\mathrm{N}_{\mathrm{0}}^{\mathrm{S}}$ in a subcritical pitchfork bifurcation $\mathrm{P}_{\mathrm{sub}}$, which destroys the superradiant equilibria and turns the South pole equilibrium unstable along branch $\mathrm{N}_{\mathrm{1}}^{\mathrm{S}}$. The following steady-state bifurcations as $\lambda_{+}$ increases feature the same structure as those in Fig. \ref{fig:SprRadBifDiag}(a).
	
	In physical terms, the key difference between the two cases, $\lambda_{-} = 1$ and $\lambda_{-} = 2$, is the manner in which the superradiant phase comes into existence.  Fig. \ref{fig:SprRadBifDiag}(a) describes the superradiant phase transition when $\lambda_{-}=1$. Here, the superradiant phase emerges with the creation of two stable superradiant equilibria, after the supercritical pitchfork bifurcation $\mathrm{P}_{\mathrm{spr}}$; this is the semiclassical description of the classic Dicke phase transition as in the case of balanced coupling $\lambda_{-}=\lambda_{+}$. However, the transition described by Fig. \ref{fig:SprRadBifDiag}(b) consists of two stages, where the emergence of the superradiant phase with definite large amplitude and the disappearance of the normal phase now occur separately. In between the two stages, the normal and superradiant phases coexist as a multistable configuration, the structure of which is discussed in Sect. \ref{Sect:Multistability}.

	\subsection*{Superradiant phase diagram}
	
\begin{figure}[h]
		\centering
		\includegraphics[width=0.48\textwidth]{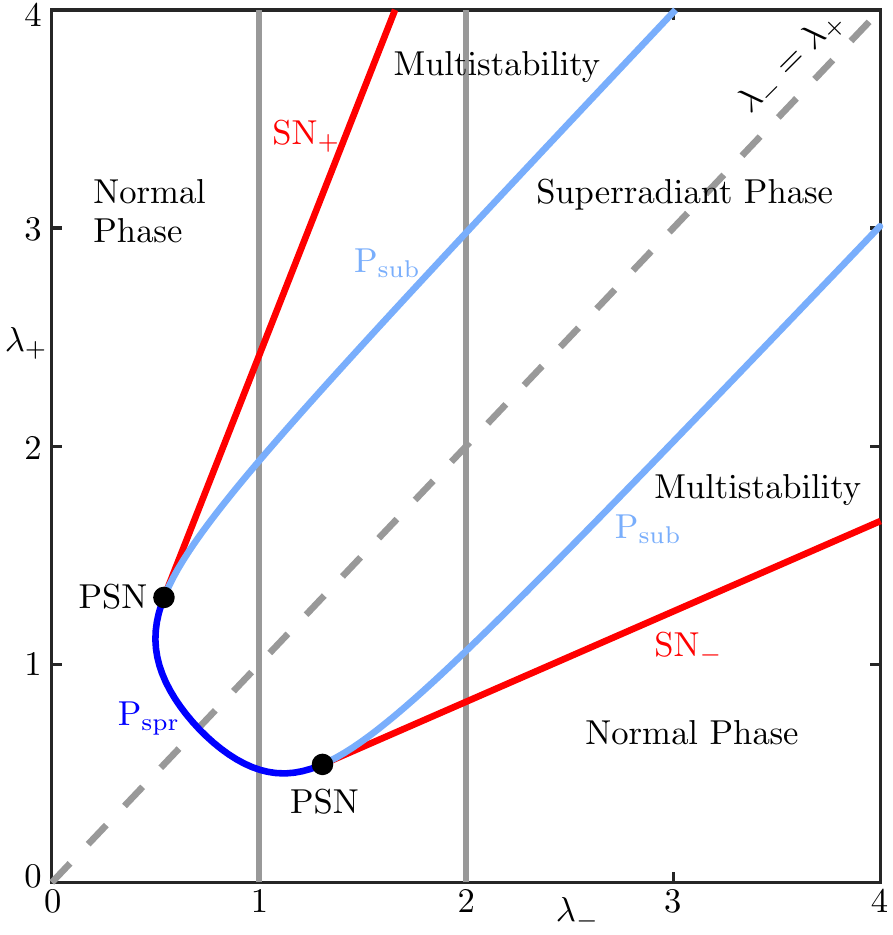}
		\caption{\label{fig:SprRad2D}Two-parameter bifurcation diagram describing the superradiant phase transition in $\lambda_{-}$ and $\lambda_{+}$. Shown are the locations of the saddle-node bifurcations SN (red curves) and the locations of the pitchfork bifurcations P (blue curve), light-blue indicates the pitchfork bifurcation is subcritical and darker blue indicates supercritical. Codimension-two pitchfork-saddle-node bifurcations are shown as black dots and labelled PSN. The vertical gray lines indicate the slices of the parameter plane that produce Figs.~\ref{fig:SprRadBifDiag}(a1) and (b1), respectively. Other parameters are set at $\kappa=\omega=\omega_0 = 1$.}
	\end{figure}

	Figure \ref{fig:SprRad2D} shows the superradiant phase diagram in the $(\lambda_{-},\lambda_{+})$-parameter plane, which outlines regions where the number of equilibria is the same. It is constructed by determining the locations of the pitchfork and saddle-node bifurcations. A necessary condition for the pitchfork and saddle-node bifurcations to occur is that the determinant of the Jacobian matrix of system (\ref{eqn:StereoEqns}), evaluated at the respective equilibrium point, is zero, i.e., $\det(J)=0$, where $J$ is the Jacobian matrix of (\ref{eqn:StereoEqns}). Evaluating at the South pole equilibrium point $(\alpha,\beta,\gamma)=(0,0,-1/2)$ yields the quartic equation
	\begin{equation}\label{eqn:PitchforkQuartic}
	(\lambda_{-}^2 - \lambda_{+}^2)^2 - 2\omega\omega_0(\lambda_{-}^2 + \lambda_{+}^2) + \omega_0^2(\kappa^2 + \omega^2) = 0,
	\end{equation}
	which is quadratic in $\lambda_{+}^2$. Its (positive) solution is
	\begin{equation}\label{eqn:Pitchfork}
	\lambda_{+} = \sqrt{\lambda_{-}^2 + \omega\omega_0 \pm \sqrt{\omega_0(4\omega\lambda_{-}^2 - \kappa^2\omega_0)}},
	\end{equation}
	which we identify as the pitchfork bifurcation curve in Fig. \ref{fig:SprRad2D} when taken as a function of $\lambda_{-}$. In particular, this curve gives the well known value $\lambda^{*}$ of the superradiant phase transition for the case of balanced coupling $\lambda^{*} = \lambda_{-} = \lambda_{+}$ \cite{dimer_proposed_2007}, namely
	\begin{equation}\label{eqn:Dimer}
	\lambda^{*} = \sqrt{\frac{\omega_0(\omega^2 + \kappa^2)}{4\omega}}.
	\end{equation}
	The asymptotic expression for the pitchfork curve in the limit of large coupling is $\lambda_{+} = \lambda_{-}\pm\sqrt{\omega\omega_0}$, which we note is independent of the cavity dissipation rate $\kappa$.
	
	Following a similar procedure, but evaluating the Jacobian matrix $J$ at the superradiant equilibrium points, we find another quartic equation that we can identify with the locations of the saddle-node bifurcations:
	\begin{equation}\label{eqn:SaddleNodeQuartic}
	\kappa^2(\lambda_{-}^2 - \lambda_{+}^2)^2 - 4\omega^2\lambda_{-}^2\lambda_{+}^2 = 0.
	\end{equation}
	The solutions to this equation give the two saddle-node bifurcation curves in the $(\lambda_{-},\lambda_{+})$-parameter plane as the lines
	\begin{equation}\label{eqn:SaddleNode}
	\lambda_{+} = \pm\left( \frac{\omega\pm\sqrt{\omega^2 + \kappa^2}}{\kappa} \right)\lambda_{-}.
	\end{equation}
	
	The pitchfork bifurcation changes criticality when the saddle-node and pitchfork bifurcation curves intersect to form codimension-two pitchfork-saddle-node bifurcations (PSN). These are, in fact, identical to the ``dissipation-induced tricritical points" studied by Soriente \emph{et al.} in \cite{soriente_dissipation-induced_2018}. However, Soriente \emph{et al.} use a different parameterization of the model (see Appendix); in our version, the PSN points are
	\begin{subequations}
		\begin{align}
		\lambda_{-}^{\mathrm{PSN}} &= \sqrt{\frac{\omega_0(\omega^2 + \kappa^2)\pm\omega\omega_0 \sqrt{\omega^2 + \kappa^2}}{2\omega}}, \\
		\lambda_{+}^{\mathrm{PSN}} &= \sqrt{\frac{\omega_0(\omega^2 + \kappa^2)\mp\omega\omega_0 \sqrt{\omega^2 + \kappa^2}}{2\omega}}.
		\end{align}
	\end{subequations}
	Multistability and a single tricritical point in the unbalanced model have also been noticed in a theoretical study of Keeling, Bhaseen, and Simons \cite{keeling_collective_2010} in a different parameter regime.
	
	In Fig. \ref{fig:SprRad2D}, there are two disconnected saddle-node curves $\mathrm{SN}_{\pm}$, defined for $\lambda_{-}\geq\lambda_{-}^{\mathrm{PSN}}$, corresponding to either the plus or minus sign in (\ref{eqn:SaddleNode}). The pitchfork bifurcation curve (\ref{eqn:Pitchfork}) is cut by the points PSN into three distinct sections where the pitchfork bifurcation is either subcritical or supercritical.

	We also see that the steady-state bifurcations are symmetric about the diagonal line $\lambda_{-}=\lambda_{+}$, i.e., invariant under the parameter exchange $\lambda_{-}\leftrightarrow\lambda_{+}$. This is due to the fact that (\ref{eqn:PitchforkQuartic}) and (\ref{eqn:SaddleNodeQuartic}) contain only even powers of $\lambda_{\pm}$.  Furthermore, in the limit of large coupling, the pitchfork bifurcation curves become parallel to the diagonal, $\lambda_{-}=\lambda_{+}$. Hence, the system cannot exhibit multistability phenomena between normal and superradiant phases in the case of balanced coupling, because the diagonal intersects the pitchfork curve only once, at the point $\lambda^{*}$ given in (\ref{eqn:Dimer}).
	
	
	\section{Multistability}\label{Sect:Multistability}
	
	As we have seen in Sect. \ref{Sect:SteadyStateBifurcations}, the generalization to unbalanced coupling in the open Dicke model allows for a case of the superradiant phase transition where the emergence of the superradiant phase and the disappearance of the normal phase occur separately. In between the two stages of this transition, the two phases coexist due to the existence of multiple stable equilibria of the system, which are ``selected" as the final state under time evolution by the initial conditions.
	
	\subsection*{Basins of attraction}
	
	The nature of the multistability can be examined by considering the basins of attraction of the normal and superradiant phase equilibria. That is, the sets of initial conditions that converge to particular stable equilibria as $t\rightarrow\infty$. In the multistable case, there exist six equilibria in total: the stable normal phase equilibrium at the South pole $\mathrm{N}_{0}^{\mathrm{S}}$, the unstable normal phase equilibrium at the North pole $\mathrm{N}_{2}^{\mathrm{N}}$, two stable superradiant equilibria $\mathrm{SR}_0$, and two saddle superradiant equilibria $\mathrm{SR_{1}}$.
	
	\begin{figure}[h!]
		\centering
		\includegraphics[width=6cm]{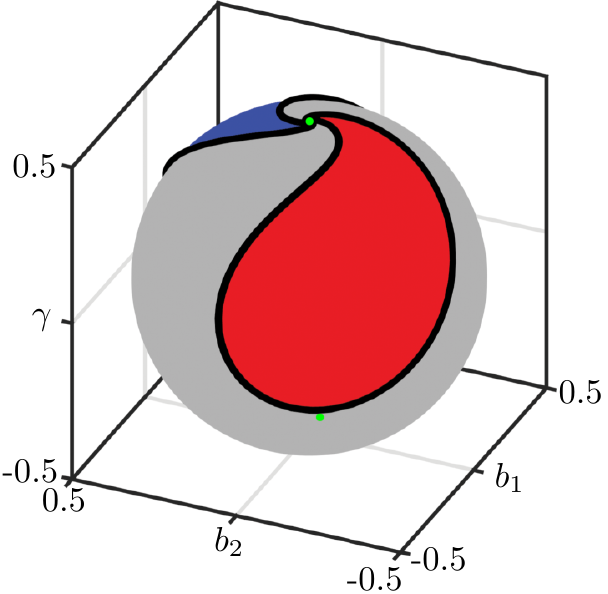}
		\caption{\label{fig:BasinOfAttraction}Basins of attraction on the Bloch sphere of the superradiant equilibria (red and blue) and the normal phase (gray). The black curve separating the basins of attraction is the stable manifold $W^{s}(\mathrm{SR}_{1})$ for $\alpha=0$; here $\lambda_{-} = 2$, $\lambda_{+} =  0.9$, $\kappa=\omega=\omega_0 = 1$.}
	\end{figure}
	
	The basins of attraction of $\mathrm{SR}_0$ and $\mathrm{N}_{0}^{\mathrm{S}}$ are separated by the \emph{stable manifolds} $W^{s}(\mathrm{SR}_1)$ of the saddle superradiant equilibria, which are the set of points in phase space that converge to these equilibria as $t\rightarrow \infty$. These equilibria have one eigenvalue of the Jacobian matrix of system (\ref{eqn:StereoEqns}) with positive real part, and three with negative real part. Thus, by the Stable Manifold Theorem \cite{guckenheimer_nonlinear_1996}, the stable manifolds are three-dimensional. To illustrate the basin, we consider initial conditions for $\alpha=0$, i.e., an initially empty cavity. In this case the boundaries of the basins of attraction are given by the one-dimensional intersection of the three-dimensional stable manifolds and the two-dimensional surface defined by $\alpha=0$. The boundaries of the basins of attraction can then be found numerically by computing the manifold $W^{s}(\mathrm{SR}_0)$ with numerical continuation techniques \cite{krauskopf_computing_2007}.
	
	In Fig. \ref{fig:BasinOfAttraction}, the basins of attraction are shown on the Bloch sphere for a parameter set in the region of multistability. The two basins of attraction of the superradiant equilibria $\mathrm{SR}_0$ appear as teardrop-shaped lobes emanating from the North pole. The region outside of the lobes is the basin of attraction for the normal phase equilibrium $\mathrm{N}_{0}^{\mathrm{S}}$.
	
	As $\lambda_{+}$ is increased, the superradiant basins of attraction grow from the North pole. They continue to grow until they reach a point where the system is ``halfway" between the superradiant and normal phases; that is, the basins of attraction occupy equal areas. The superradiant basins then continue to grow until they entirely envelop the Bloch sphere at the second pitchfork bifurcations, after which the system is in the superradiant phase.
	
	
	\section{Emergence of oscillations}\label{Sect:Oscillations}
	
	\begin{figure}[h!]
		\centering
		\includegraphics[width=8.6cm]{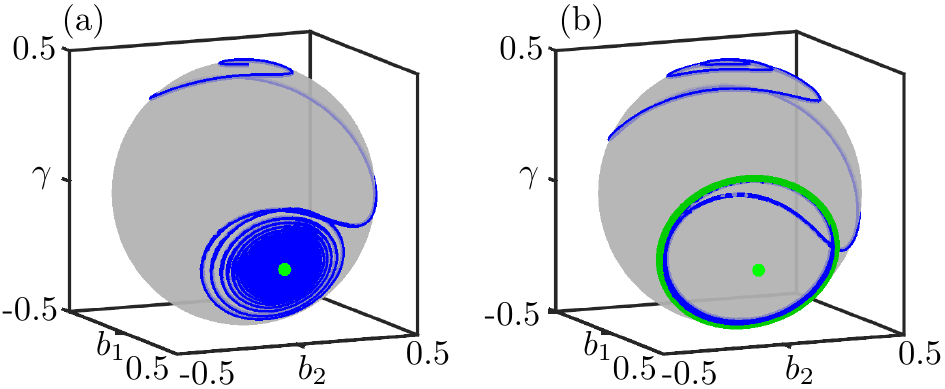}
		\caption{\label{fig:HopfBifExample}Creation of a periodic orbit on the Bloch sphere in a Hopf bifurcation, (a) $\lambda_{+}=1.2$, (b) $\lambda_{+}=1.4$, with $\kappa=\omega=\omega_0 = 1$; shown is a trajectory from an initial condition near the North pole with zero photon number.}
	\end{figure}
	
	Persistent oscillations in the semiclassical model arise due to a pair of Hopf bifurcations, where a stable periodic orbit bifurcates from a superradiant equilibrium point, after which the superradiant equilibrium point becomes unstable. After the Hopf bifurcation, two asymmetric periodic orbits $\Gamma$ emerge after the transition due to $\mathbb{Z}_{2}$-symmetry. Figure \ref{fig:HopfBifExample} shows this on the Bloch sphere. Before the bifurcation, in Fig. \ref{fig:HopfBifExample}(a), the system is in the superradiant phase and trajectories converge to the superradiant equilibria $\mathrm{SR}_0$. After the Hopf bifurcation, shown in Fig. \ref{fig:HopfBifExample}(b), the stable periodic orbit $\Gamma$ emerges from the (now unstable) superradiant equilibrium point; this corresponds to the oscillatory phase as observed experimentally in \cite{zhiqiang_nonequilibrium_2017}. Note that a self-intersection on the Bloch sphere in Fig. \ref{fig:HopfBifExample} is not a violation of the uniqueness theorem as the Bloch sphere is a projection of the state space.
	
\begin{figure}[t]
		\centering
		\includegraphics{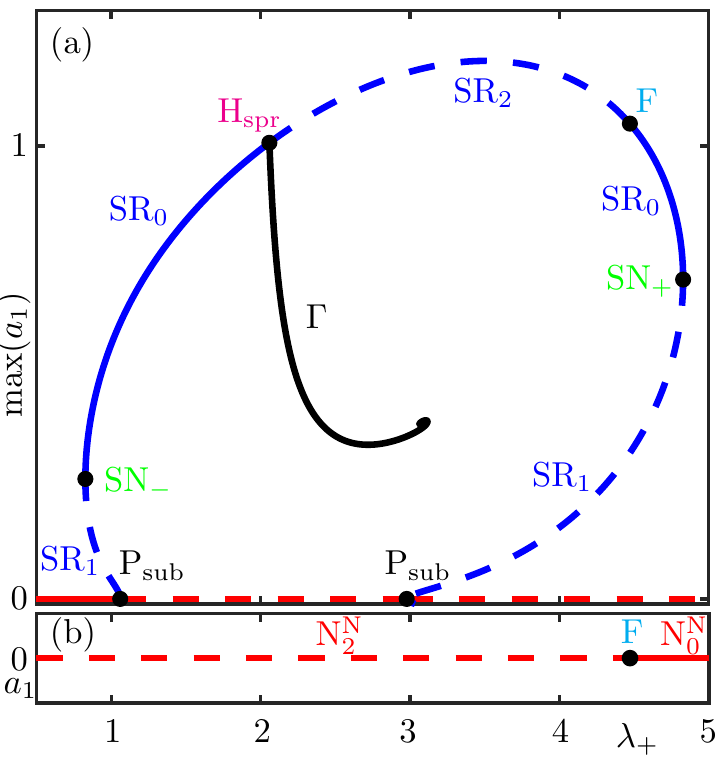}
		\caption{\label{fig:HopfBifDiag}One-parameter bifurcation diagram in panel (a) for fixed $\lambda_{-}=2$, $\kappa=\omega_0=\omega=1$ with the branch (black) of periodic orbits $\Gamma$ emerging from the supercritical Hopf bifurcation $\color{black}{\mathrm{H}_{\mathrm{spr}}}$. Panel (b) shows the bifurcation diagram of the North pole; compare with Fig. \ref{fig:SprRadBifDiag}.}
	\end{figure}

	The bifurcation diagram for $\lambda_{-}=2$ in Fig. \ref{fig:HopfBifDiag}(a) shows the branch of the bifurcating periodic orbits as represented by their maximum amplitude in $a_1$. The pair of periodic orbits $\Gamma$ emerge in a pair of supercritical Hopf bifurcations $\color{black}{\mathrm{H}_{\mathrm{spr}}}$ after the first saddle-node bifurcation $\color{black}{\mathrm{SN}}_{-}$. The periodic orbit $\Gamma$ disappears near $\lambda_{+}\approx 3.1$ in a global bifurcation, which will be discussed in Sect. \ref{Sect:Chaos}.
	
	
	\begin{figure}[h]
		\centering
		\includegraphics[width=8.6cm]{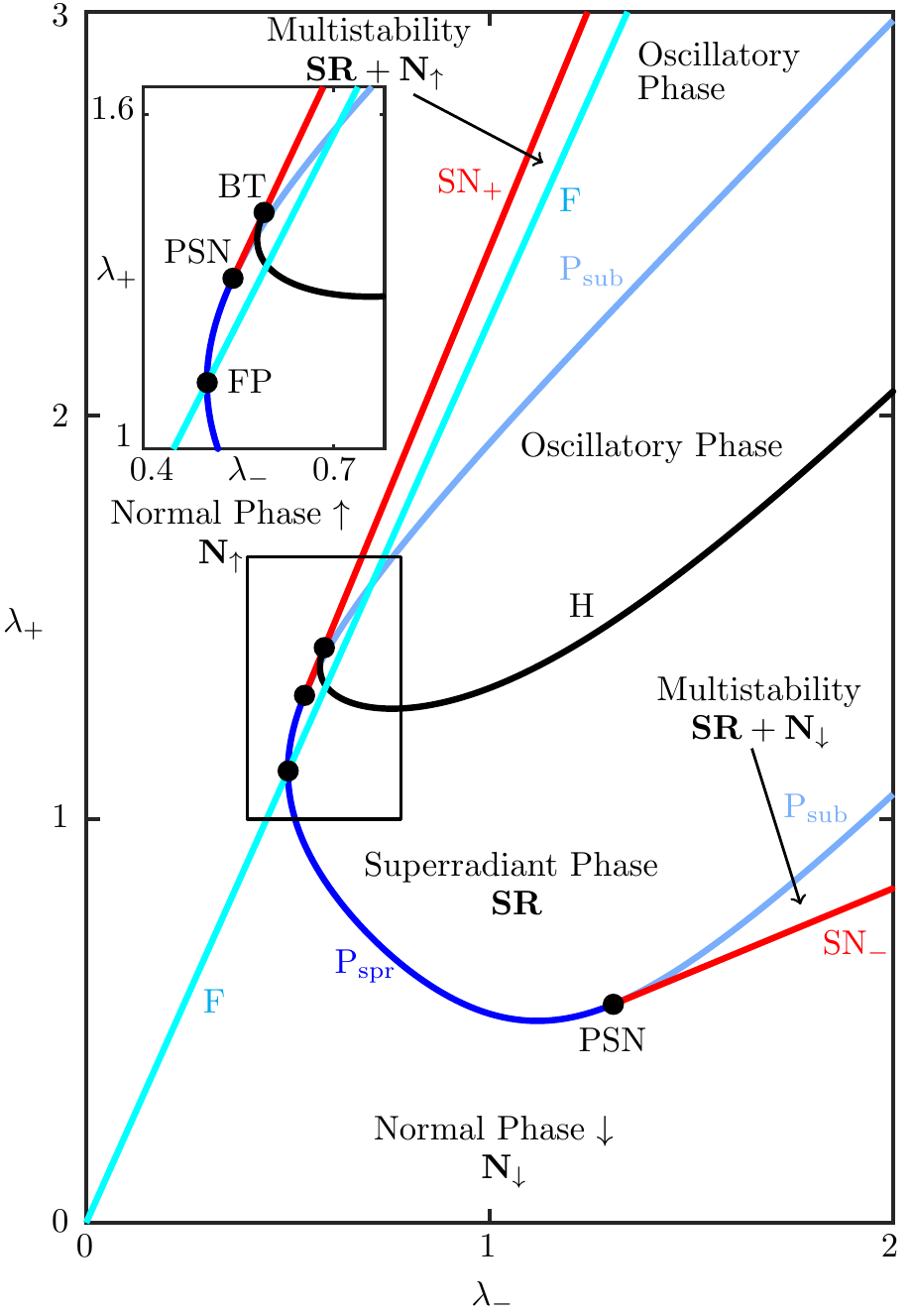}
		\caption{\label{fig:Hopf2D}The two-parameter bifurcation diagram describes the oscillatory phase transition in $\lambda_{-}$ and $\lambda_{+}$. Hopf bifurcations are located on the black curve H. The oscillatory region is bounded by the curves H and F; compare with Fig. \ref{fig:SprRad2D}. Here $\kappa=\omega=\omega_0=1$.}
	\end{figure}
	As with the superradiant transition, we map the oscillatory phase transition with respect to both coupling strengths by continuing the Hopf bifurcation in the $(\lambda_{-},\lambda_{+})$-parameter plane. The oscillatory phase diagram is presented in Fig. \ref{fig:Hopf2D}, where the Hopf bifurcation curve is labelled H. The oscillatory phase is located in the region bounded by the Hopf bifurcation curve H and the curve F. As demonstrated in Fig. \ref{fig:SprRadBifDiag}, the curve F marks a change in stability of the North pole equilibrium point ($\mathrm{N}^{\mathrm{N}}_{2}\rightarrow\mathrm{N}^{\mathrm{N}}_{0}$), we therefore refer to this transition as the \emph{pole-flip transition}, to be discussed in Sect. \ref{Sect:PoleFlipTransition}. The inclusion of this curve in the oscillatory phase diagram Fig. \ref{fig:Hopf2D} allows us to distinguish regions of the two types of normal phases: namely, spin-up where the North pole is stable $\mathrm{N}^{\mathrm{N}}_0$ and spin-down where the South pole is stable $\mathrm{N}^{\mathrm{S}}_0$. The spin-down and spin-up normal phase regions are denoted $\mathbf{N}_{\downarrow}$ and $\mathbf{N}_{\uparrow}$, respectively. The superradiant region is denoted $\mathbf{SR}$ and the multistable regions are denoted $\mathbf{SR}+\mathbf{N}_{\downarrow}$ and $\mathbf{SR}+\mathbf{N}_{\uparrow}$.
	
	The inset of Fig. \ref{fig:Hopf2D} gives details of two additional bifurcations that arise due to oscillations in system (\ref{eqn:StereoEqns}). The first is a codimension-two Bogdanov-Takens bifurcation (BT) where the Hopf bifurcation curve ends on the saddle-node curve $\mathrm{SN}_{+}$  \cite{guckenheimer_nonlinear_1996,kuznetsov_elements_2004}. Here, the Jacobian matrix $J$ features two zero-eigenvalues and satisfies $\det(J) = \mathrm{Tr}(J) = 0$. The second codimension-two point, which we label FP, is the first intersection of the pole-flip transition curve F with the pitchfork curve $\mathrm{P}_{\mathrm{spr}}$.
	
	
	\section{Pole-Flip Transition}\label{Sect:PoleFlipTransition}
	
	Figure \ref{fig:Hopf2D} shows the pole-flip transition curve F that forms the boundary of the spin-down normal phase ($\gamma\rightarrow-1/2$), where the South pole of the Bloch sphere is stable, and the spin-up normal phase ($\gamma\rightarrow1/2$), where the North pole is stable. This transition therefore describes quite a dramatic change to the atomic dynamics, as the steady-state spontaneously changes from all atoms occupying their ground states, to all atoms becoming excited.
	
\begin{figure}[h]
		\centering
		\includegraphics[width=8.6cm]{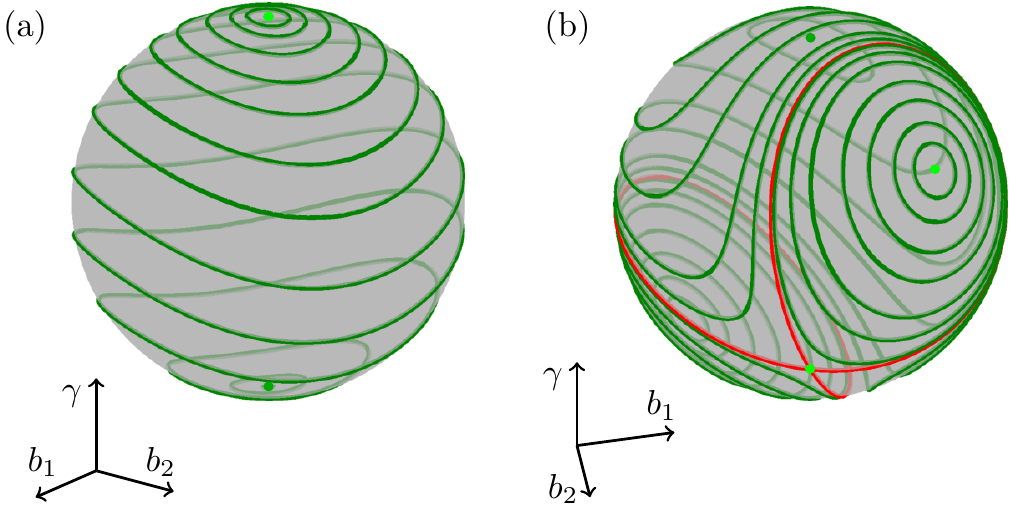}
		\caption{\label{fig:PoleFlip}Foliations of periodic orbits on the Bloch sphere during the pole-flip transition: (a) normal pole-flip transition for $\lambda_{-}=2/5$, $\lambda_{+}=2/\sqrt{5}$, (b) superradiant pole-flip transition for $\lambda_{-}=1$, $\lambda_{+}=\sqrt{5}$.}
	\end{figure}

	Along the pole-flip transition curve F, the Bloch sphere is foliated by an infinite number of periodic orbits, see Fig. \ref{fig:PoleFlip}. The North pole equilibrium point also features two eigenvalues with zero real part. To solve for the curve F, we use the bialternate product $\odot$ \cite{kuznetsov_elements_2004}. It has the property that for a matrix $A$, the matrix $A\odot 2I$ has eigenvalues that are pairwise sums of the eigenvalues of $A$ \cite{stephanos_sur_1900}, where $I$ is the identity matrix. Parameter values where the Jacobian matrix $J$ has eigenvalues with zero real part can then be found when the matrix $J\odot 2I$ has a zero eigenvalue, i.e., solving $\det(J\odot 2I)=0$. Evaluating this determinant at the North pole equilibrium point gives us the analytical expression for the curve F
	\begin{equation}\label{eqn:PoleFlip}
	\lambda_{+} = \sqrt{\frac{\kappa^2 + (\omega + \omega_0)^2}{\kappa^2 + (\omega - \omega_0)^2}}\ \lambda_-.
	\end{equation}
	Similarly, the determinant could be used to compute Hopf bifurcations of the superradiant equilibria; however, this requires analytic expressions for these equilibria, which are quite complicated and not particularly enlightening. The parameter values for the points where the pitchfork bifurcation and pole-flip transition occur simultaneously, on the other hand, are readily found from (\ref{eqn:PoleFlip}) and (\ref{eqn:Pitchfork}) as
	\begin{subequations}
		\begin{align}
		\lambda_{\pm}^{\mathrm{FP}_1} &= \frac{1}{2}\sqrt{\frac{\omega_0(\kappa^2+(\omega\pm\omega_0)^2)}{\omega}}, \\
		\lambda_{\pm}^{\mathrm{FP}_2} &= \frac{1}{2}\sqrt{\frac{(\kappa^2+\omega^2)(\kappa^2+(\omega\pm\omega_0)^2)}{\omega\omega_0}}.
		\end{align}
	\end{subequations}
	
	There are two cases corresponding to two different foliations of the Bloch sphere. The first case consists of periodic orbits that all oscillate about the poles, which is illustrated in Fig. \ref{fig:PoleFlip}(a). This is the boundary case of the two configurations of the normal phase, spin-up or spin-down, and is located along the curve F for $\lambda_{-}<\min(\lambda_{-}^{\mathrm{FP}_1},\lambda_{-}^{\mathrm{FP}_2})$. We refer to this case as the \emph{normal pole-flip transition}. The second case occurs after the pitchfork bifurcation creates a pair of superradiant equilibria. Here oscillations develop around these equilibria, confined by a pair of orbits homoclinic to the South pole, forming a figure eight, as illustrated in Fig. \ref{fig:PoleFlip}(b). This transition, which we refer to as the \emph{superradiant pole-flip transition}, occurs along the curve F when $\lambda_{-}>\min(\lambda_{-}^{\mathrm{FP}_1},\lambda_{-}^{\mathrm{FP}_2})$.
	
	At the point of the pole-flip transition, there is a two-dimensional surface of periodic orbits which features a continuous symmetry, and thus has an associated conserved quantity. We find that this conserved quantity is the semiclassical energy per atom, given by $E=\langle \hat{H} \rangle/N$, that is,
	\begin{equation}
	E = \omega(a_{1}^2 + a_{2}^2) + \omega_0\gamma + 2(\lambda_{+}+\lambda_{-})a_{1}b_{1} - 2(\lambda_{+}-\lambda_{-})a_{2}b_{2}
	\end{equation}
	is conserved along the two-dimensional surface foliated by periodic orbits. This property was confirmed by careful numerical investigation by calculating the energy $E$ along a selection of periodic orbits computed via numerical continuation. 
	
	Moreover, this surface of periodic orbits is attracting, meaning orbits with initial conditions that lie outside of the surface will converge to a periodic orbit with constant energy. Hence, generically, orbits will initialize outside the surface of periodic orbits and during their convergence their energy will change until the energy approaches the constant energy of a particular periodic orbit, not necessarily with the same initial energy.

	The presence of energy-conserving periodic orbits in an open system such as this is surprising. These periodic orbits represent the balance of processes that preferentially drive the spins up or down, where there is also a balance of the energy exchange between the atoms and the cavity field, even in the presence of dissipation. However, as shown in a recent article \cite{shchadilova_fermionic_2020}, this behavior is modified significantly with the addition of spin dissipation and pumping.
	

	\section{Emergence of chaotic dynamics}\label{Sect:Chaos}
	
	We now turn our attention to describing and analyzing the bifurcations of periodic orbits, in particular, those that lead to the onset of chaos. We observe period-doubling cascades leading to the formation of two chaotic attractors, related by mirror symmetry, followed by their collision to form a single chaotic attractor through a sequence of global bifurcations.
	
	\subsection{Period-doubling cascades}
	
	In a period-doubling bifurcation, a periodic orbit bifurcates to form a new periodic orbit with twice the period of the original, effectively modulating the amplitude of oscillation with a period of two.

	\begin{figure}[h]
		\centering
		\includegraphics[width=8.6cm]{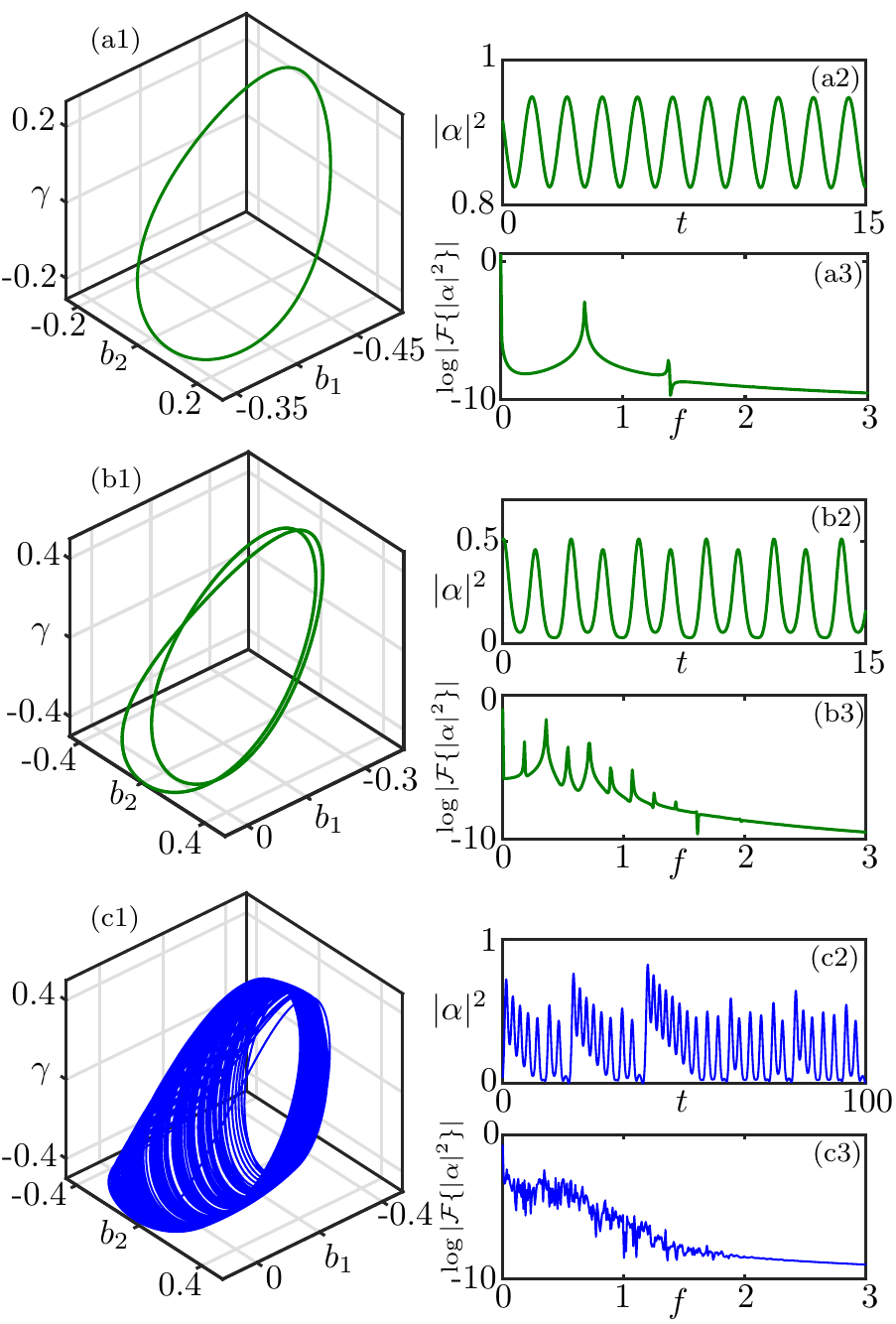}
		\caption{\label{fig:PeriodDoubling}Period-doubling cascade on the Bloch sphere (a1-c1), in the photon number temporal traces (a2-c2), and in the Fourier spectra (a3-c3), for (a) $\lambda_{+}=1.67$, (b) $\lambda_{+}=2.13$, (c) $\lambda_{+}=2.225$; other parameters are $\kappa = \omega = \omega_0 = 1$, $\lambda_{-}=1.5$.}
	\end{figure}
	
	A very common route to chaos is through a sequence or cascade of period-doubling bifurcations leading to the creation of chaotic attractors \cite{guckenheimer_nonlinear_1996}. Note that due to the system's $\mathbb{Z}_2$-symmetry, most periodic orbits and chaotic attractors have symmetric counterparts; hence, period-doubling bifurcations also occur in pairs.
	
	A period-doubling route to chaos, shown in Fig. \ref{fig:PeriodDoubling}, can be observed in system (\ref{eqn:StereoEqns}). Here, we show trajectories for three values of $\lambda_{+}$ (with $\lambda_{-}$ held constant) on the Bloch sphere, temporal traces and Fourier spectra $\log|\mathcal{F}\{|\alpha|^2\}|$ of the photon number (where $\mathcal{F}\{\cdot\}$ denotes the Fourier transform). The system is in the oscillatory phase in Fig. \ref{fig:PeriodDoubling}(a): here the two periodic orbits (only one of which is shown) feature a single loop [Fig. \ref{fig:PeriodDoubling}(a1)] and the photon number $|\alpha|^2$ temporal trace is shown in Fig. \ref{fig:PeriodDoubling}(a2). Furthermore, the Fourier spectrum has a single prominent peak at the frequency of oscillation [Fig. \ref{fig:PeriodDoubling}(a3)], and a much smaller first harmonic. Upon increase of $\lambda_{+}$ (as in Fig. \ref{fig:PeriodDoubling}(b)), a period-doubling bifurcation has occurred; the original periodic orbit on the Bloch sphere is now unstable (and not shown) and a periodic orbit with two loops is now the attractor [Fig. \ref{fig:PeriodDoubling}(b1)]. The temporal trace of the photon number shows the amplitude modulation of the oscillation cycle [Fig. \ref{fig:PeriodDoubling}(b2)], and the Fourier spectrum shows the emergence of peaks in between the previous peaks [Fig. \ref{fig:PeriodDoubling}(b3)], corresponding to additional contributions from other oscillation frequencies. As $\lambda_{+}$ is increased further (as in Fig. \ref{fig:PeriodDoubling}(c)), the periodic orbits bifurcate many more times as the system undergoes an infinite sequence of period-doubling bifurcations and two chaotic attractors emerge on the Bloch sphere (Fig. \ref{fig:PeriodDoubling}(c1) shows one of them). The temporal trace of the photon number shows a more complicated, non-repeating oscillation cycle [Fig. \ref{fig:PeriodDoubling}(c2)], and the Fourier spectrum is now broad with many peaks due to contributions from infinitely many oscillation frequencies [Fig. \ref{fig:PeriodDoubling}(c3)], which is characteristic of chaotic motion \cite{nayfeh_applied_1995}.

	\begin{figure}[h]
		\centering
		\includegraphics[width=8.6cm]{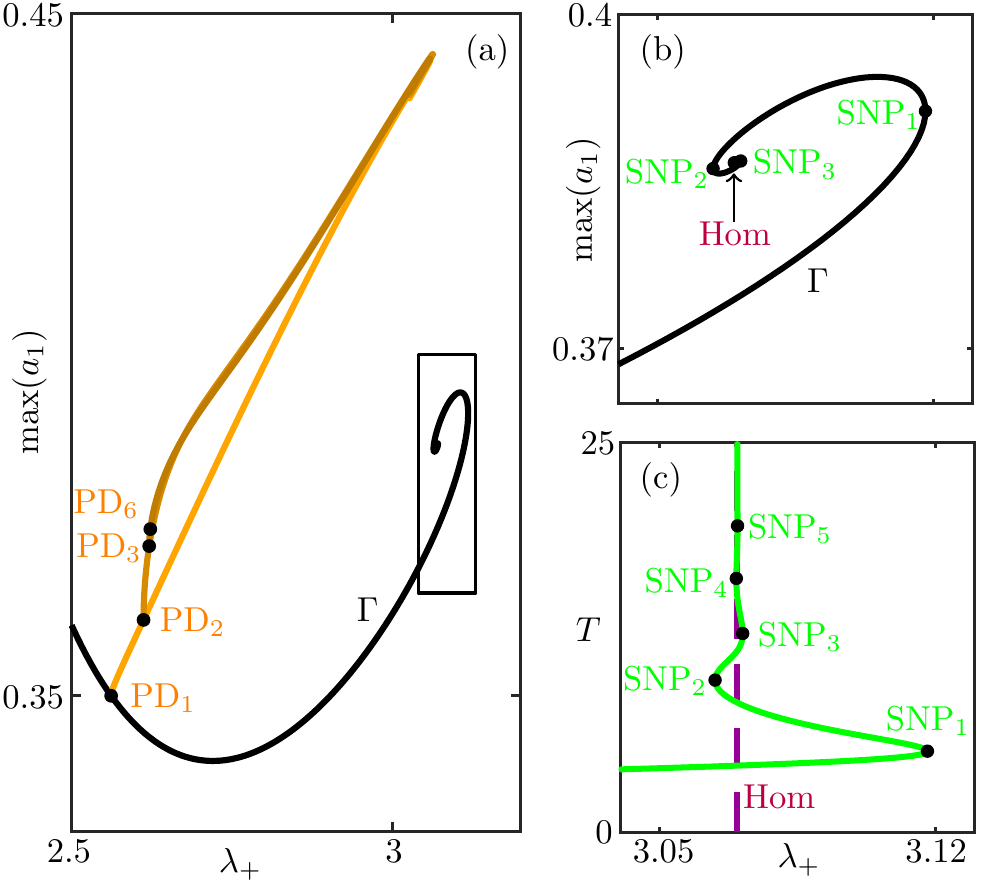}
		\caption{\label{fig:PDBifDiag}One-parameter bifurcation diagram (a) showing the primary (black curve) and period-doubled (orange curves) branches of periodic orbits; the inset shows the period $T$ of the first period-doubled orbit. Panel (b) shows an enlargement of the primary branch as a homoclinic bifurcation denoted Hom is approached. The spiral structure of the primary branch is shown in panel (c) in terms of the period $T$ of the periodic orbit. Here $\kappa = \omega = \omega_0 = 1$, $\lambda_{-}=2$.}
	\end{figure}
	
	Figure \ref{fig:PDBifDiag} shows a one-parameter bifurcation diagram for $\lambda_{-}=2$ and varying $\lambda_{+}$; here, we represent periodic solutions by the maximum of the real part of the cavity field amplitude in panels (a) and (b). We observe a cascade of period-doubling bifurcations (PD) bifurcating off of the primary branch of periodic orbits $\Gamma$. Every period-doubled branch of periodic orbits then has another period-doubling bifurcation along it, where another branch of periodic orbits emerges. The accumulation point of the period-doubling cascade is approximated near $\mathrm{PD}_{6}$, which occurs at $\lambda_{+}\approx 2.624$.

	\subsection{Global bifurcations to chaos}
	
	\begin{figure}[h]
		\centering
		\includegraphics[width=8.6cm]{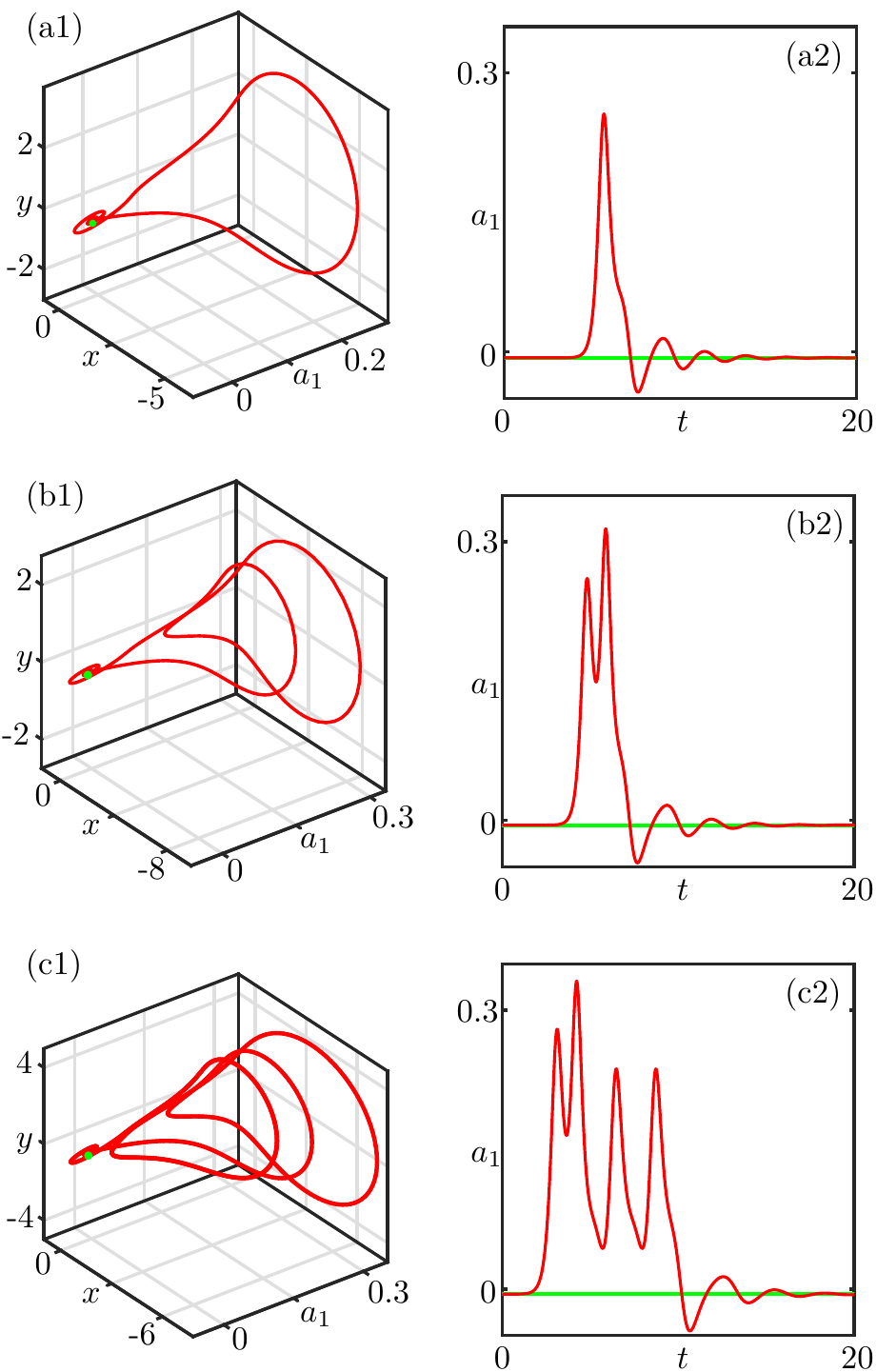}
		\caption{\label{fig:Shilnikov}Shilnikov homoclinic orbits arising from saddle-focus equilibria in $(a_{1},x,y)$-space (a1-c1) and temporal traces of the real part of the cavity field amplitude $a_{1}$ (a2-c2). Panels (a)-(c) show a 1-homoclinic orbit for $(\lambda_{-},\lambda_{+})=(5.75,7.05)$, a 2-homoclinic orbit for $(\lambda_{-},\lambda_{+})=(5.74,7.00)$, and a 4-homoclinic orbit for $(\lambda_{-},\lambda_{+})=(5.01,6.22)$, respectively. Here $\kappa = \omega = \omega_0 = 1$.}
	\end{figure}
	Numerical continuation of the periodic orbits emanating from Hopf bifurcations (see Fig. \ref{fig:HopfBifDiag}) reveals an interesting spiral structure of the primary branch of periodic orbits $\Gamma$, shown in Fig. \ref{fig:PDBifDiag}(a), with an enlarged version in Fig. \ref{fig:PDBifDiag}(b). The turning points (with respect to $\lambda_{+}$) of the spiral are \emph{saddle-node bifurcations of periodic orbits}, denoted SNP \cite{guckenheimer_nonlinear_1996}, where a pair of stable and unstable periodic orbits collide and vanish, or are created. In the case observed in Fig. \ref{fig:PDBifDiag}(b), there are sequences of saddle-node bifurcations of periodic orbits accumulating onto a bifurcation point Hom. This oscillating behavior is more apparent in Fig. \ref{fig:PDBifDiag}(c), where the bifurcation diagram of the period $T$ is shown. Here, saddle-node bifurcations of periodic orbits SNP occur at the turning points or folds of the curve. Notice that the period $T$ of the periodic orbit increases without bound, and accumulates to a value of $\lambda_{+}$ where system (\ref{eqn:SemiclassicalEqns}) exhibits a homoclinic bifurcation Hom. Here, system (\ref{eqn:SemiclassicalEqns}) exhibits a pair of trajectories that converge forward and backward in time to a pair of saddle equilibria, forming two \emph{homoclinic orbits}, in this case, to the superradiant equilibria $\mathrm{SR}_1$. The following eigenvalue inequality at the moment of the homoclinic bifurcation Hom is satisfied
	\begin{equation}\label{eqn:ShilnikovTheorem}
	\frac{-\mathrm{Re}(v_{s})}{v_{u}} < 1,
	\end{equation}
	where $v_{s} \approx -0.469\pm2.729i$ and $v_u \approx 4.318$ are the (stable) complex and (unstable) leading eigenvalues of the superradiant equilibria. Hence, the homoclinic bifurcation point (Hom) that is exhibited in system (\ref{eqn:SemiclassicalEqns}) corresponds to a wild Shilnikov bifurcation \cite{homburg_homoclinic_2010}; and it is responsible for the existence of infinitely many saddle-node bifurcations and the oscillatory behavior observed in Fig.~\ref{fig:PDBifDiag}. The Shilnikov theorem \cite{shilnikov_case_1965,shilnikov_contribution_1970} then asserts the existence of infinitely many $m$-homoclinic bifurcations, where the associated homoclinic orbit makes $m$ loops around the primary homoclinic orbit.
	
	Figure \ref{fig:Shilnikov}(a1) shows the Shilnikov homoclinic orbit in $(a_1,x,y)$-space, as computed by numerical continuation of the primary periodic orbits up to a sufficiently high period. The temporal trace of $a_{1}$ along the orbit is shown in Fig. \ref{fig:Shilnikov}(a2); note the Shilnikov orbit's characteristic tail of decaying oscillations.
	
	Additionally, we show in Fig. \ref{fig:Shilnikov}(b-c) 2-homoclinic and 4-homoclinic orbits, respectively, out of the infinitely many that exist nearby the point Hom. These homoclinic orbits are computed, also by continuation, as period-doubled orbits of sufficiently high period. The temporal traces of $a_{1}$ show the formation of additional peaks corresponding to each loop of the homoclinic orbits at the moment of the corresponding 2-homoclinic and 4-homoclinic bifurcation, shown in Fig. \ref{fig:Shilnikov}(b2) and (c2).
	
	\begin{figure}[h]
		\centering
		\includegraphics[width=8.6cm]{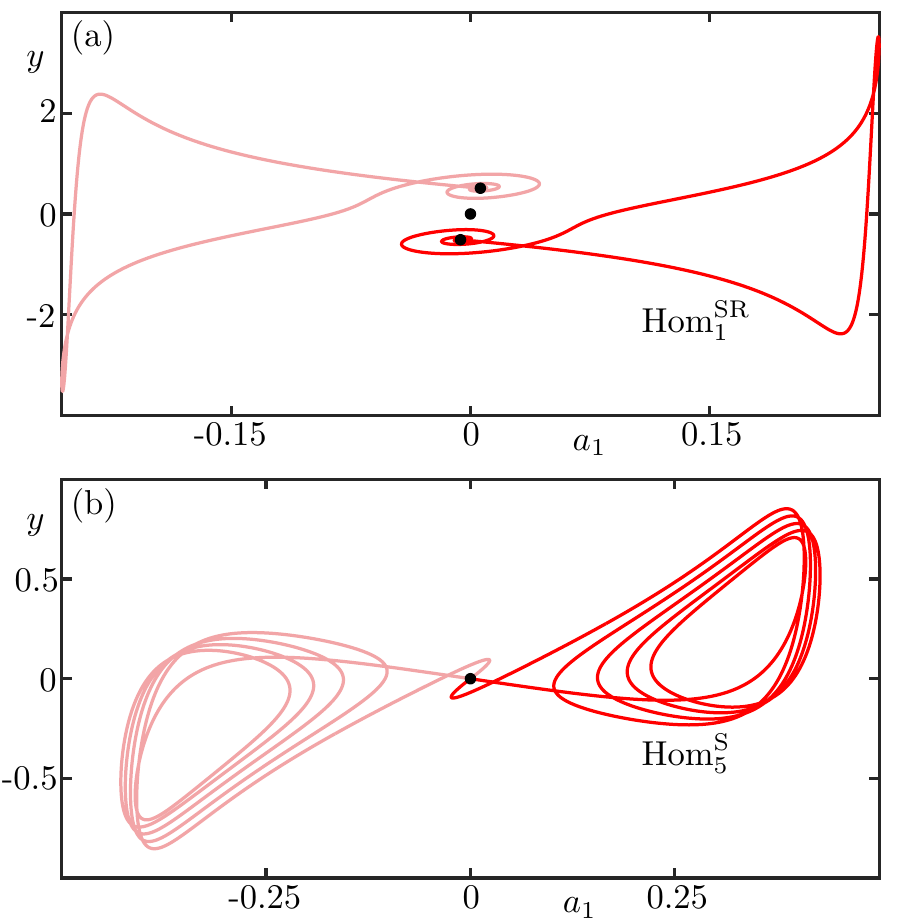}
		\caption{\label{fig:HomoclinicCompare}Two types of homoclinic orbits found in the system (\ref{eqn:StereoEqns}). Panel (a) shows a pair of homoclinic orbits to the two (saddle-focus) superradiant equilibria $\mathrm{SR}_1$ for $(\lambda_{-},\lambda_{+})=(5.75,7.05)$. Panel (b) shows a pair of homoclinic orbits to the (real saddle) normal phase equilibrium $\mathrm{N}_{1}^{\mathrm{S}}$ for $(\lambda_{-},\lambda_{+})=(0.8,1.627)$. Here $\kappa = \omega = \omega_0 = 1$.}
	\end{figure}
	
	Figure~\ref{fig:HomoclinicCompare}(a) illustrates in the $(a_{1},y)$-plane the Shilnikov homoclinic orbits that were shown in Fig.~\ref{fig:Shilnikov}(a). Here, we show both asymmetric homoclinic orbits; these are homoclinic to superradiant equilibria, so are labelled $\mathrm{Hom}_{1}^{\mathrm{SR}}$ with the subscript referring to the number of their loops. In addition to Shilnikov homoclinic orbits to asymmetric equilibria, we also find (at different parameter values) a pair of homoclinic orbits to the South pole of the Bloch sphere $\mathrm{N}_{1}^{\mathrm{S}}$, which is the origin of the stereographically projected system~(\ref{eqn:StereoEqns}); these homoclinic orbits are denoted $\mathrm{Hom}^{\mathrm{S}}_{i}$ and Fig. \ref{fig:HomoclinicCompare}(b) shows the pair $\mathrm{Hom}_{5}^{\mathrm{S}}$ that make five loops and connect to the saddle equilibrium $\mathrm{N}_{1}^{\mathrm{S}}$ on the symmetry subspace of the parity transformation (\ref{eqn:StereoParityTrans}). Furthermore, $\mathrm{N}_{1}^{\mathrm{S}}$ has real eigenvalues and, hence, the pair of homoclinic orbits feature exponentially decaying tails without oscillations.
	
\begin{figure}[h]
		\centering
		\includegraphics[width=8.6cm]{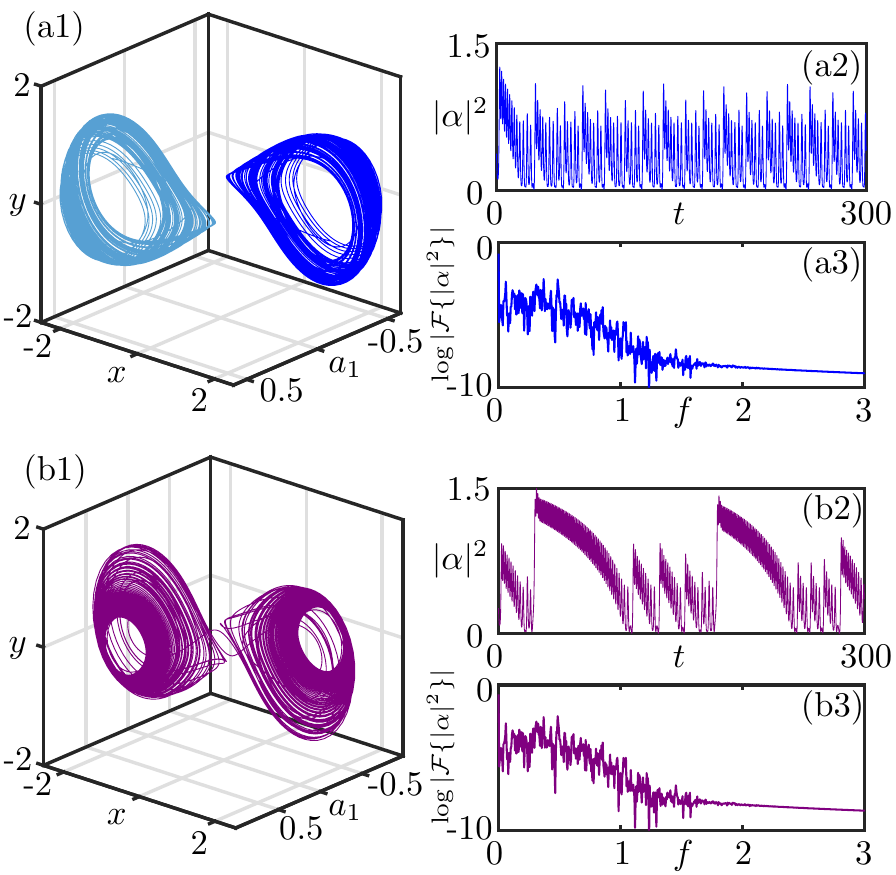}
		\caption{\label{fig:ChaosCollide}Collision of a pair of asymmetric chaotic attractors to form a single symmetric chaotic attractor. For $\lambda_{+}=2.21$ in (a1), there are two isolated chaotic attractors, shown in light and dark blue. The temporal trace of the photon number is given in (a2), and the Fourier spectrum of the photon number is shown in (a3). For $\lambda_{+}=2.26$, the merged, symmetric chaotic attractor is shown in (b1), with the temporal trace of the photon number in (b2), and the Fourier spectrum in (b3).}
	\end{figure}

	Nearby (in parameter space) the homoclinic bifurcation to the South pole equilibrium $\mathrm{N}_{1}^{\mathrm{S}}$, we find an interesting phenomenon related to the two chaotic attractors formed via period-doubling cascades and isolated in phase space, shown before in Fig. \ref{fig:PeriodDoubling}(c). We observe in numerical simulations the formation of a single, larger chaotic attractor as parameters are varied. Figure \ref{fig:ChaosCollide}(a1) shows the two isolated chaotic attractors in $(a_{1},x,y)$-space, with the temporal trace of the photon number given in Fig. \ref{fig:ChaosCollide}(a2). The two separated chaotic attractors then merge into a single chaotic attractor, illustrated in Fig. \ref{fig:ChaosCollide}(b1). This transition connects two previously isolated regions of phase space, for positive and negative $a_{1}$, and produces a dramatic change in the dynamics of the system. The most obvious change is noticeable in the temporal trace of the photon number, shown in Fig. \ref{fig:ChaosCollide}(b2), which now displays large jumps corresponding to when the state switches from one region of phase space to the other. This difference in the photon number dynamics should, in principle, be distinguishable experimentally, provided the timescale on which the experiments take place can be resolved. The Fourier spectra before and after the collision of chaotic attractors are given in Fig. \ref{fig:ChaosCollide}(a3-b3), respectively. Notice that there is no appreciable difference in the spectrum before and after the collision.
	\begin{figure}
	\centering
	\includegraphics[width=8cm]{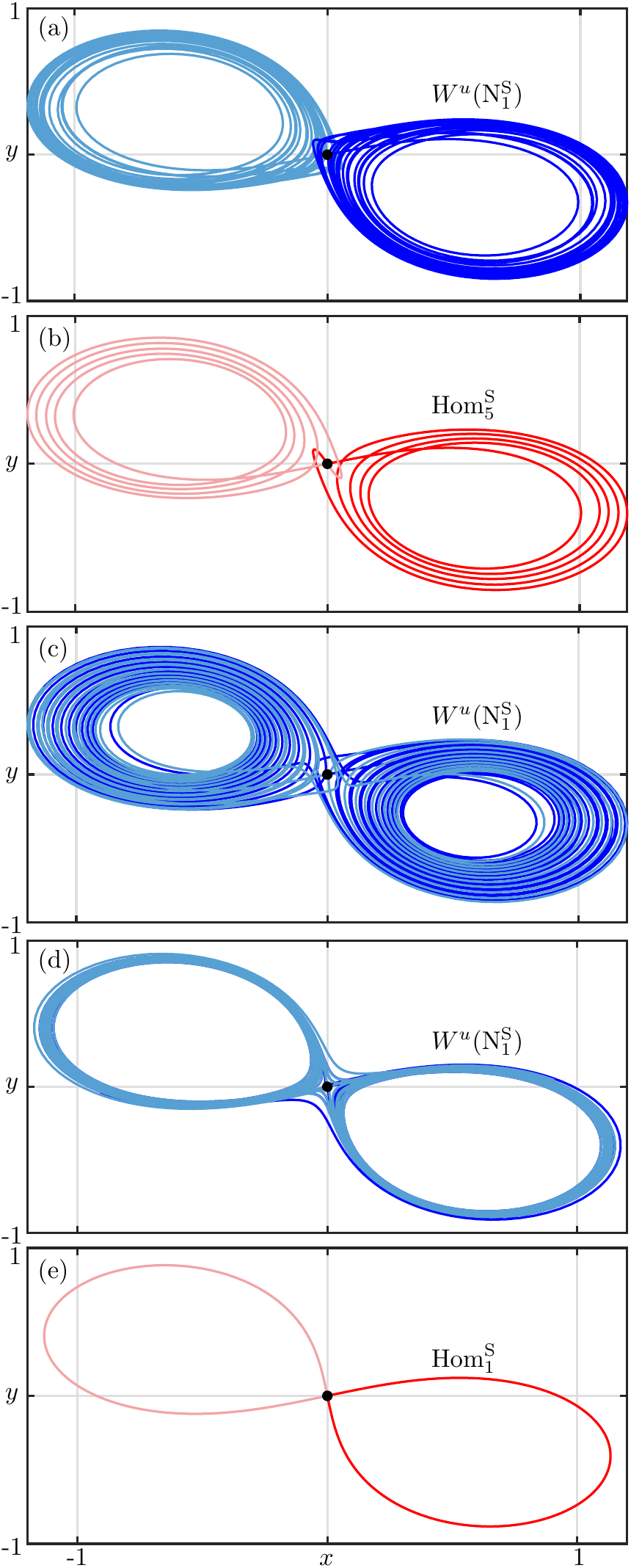}
	\caption{\label{fig:HomoclinicCollide}Unstable manifold $W^{u}(\mathrm{N}_{1}^{\mathrm{S}})$ of the origin $\mathrm{N}_{1}^{\mathrm{S}}$ of the stereographically projected system (\ref{eqn:StereoEqns}), shown in the $(x,y)$-plane; the left branch is shown in a light shade and the right darker. The unstable manifold $W^{u}(\mathrm{N}_{1}^{\mathrm{S}})$ is shown in blue, except at a homoclinic bifurcation where it is shown in red. For all panels $\kappa=\omega=\omega_0=1$, $\lambda_{-}=0.8$. Panel (a) is for $\lambda_{+}=1.61$, panel (b) for $\lambda_{+}=1.627$, panel (c) for $\lambda_{+}=1.63$, panel (d) for $\lambda_{+}=1.78$, and panel (e) for $\lambda_{+}=4/\sqrt{5}$.}
\end{figure}
	
	The collision of asymmetric chaotic attractors shown in Fig. \ref{fig:ChaosCollide} must occur through interaction with the symmetric subspace of the parity transformation (\ref{eqn:StereoParityTrans}). We find that this global transition involves homoclinic orbits of the symmetric equilibrium $\mathrm{N}_{1}^{\mathrm{S}}$, that is, the South pole of the Bloch sphere. Each panel of Fig. \ref{fig:HomoclinicCollide} shows the one-dimensional unstable manifold $W^{u}(\mathrm{N}_{1}^{\mathrm{S}})$ of $\mathrm{N}_{1}^{\mathrm{S}}$. When the two chaotic attractors are separated as in Fig. \ref{fig:HomoclinicCollide}(a), each side of the manifold $W^{u}(\mathrm{N}_{1}^{\mathrm{S}})$ accumulates onto each respective localized chaotic attractor. Figure \ref{fig:HomoclinicCollide}(b) shows the homoclinic orbit $\mathrm{Hom}_{5}^{\mathrm{S}}$, which exists for $\lambda_{+}\approx 1.627$, connects the two regions of phase space. In Fig. \ref{fig:HomoclinicCollide}(c), where $\lambda_{+}$ has been increased further, the unstable manifold $W^{u}(\mathrm{N}_{1}^{\mathrm{S}})$ accumulates onto the merged chaotic attractor; note in particular, that each side of the unstable manifold $W^{u}(\mathrm{N}_{1}^{\mathrm{S}})$ accumulates on the entire, non-localized attractor. 
	
	As $\lambda_{+}$ is increased further, the merged chaotic attractor contracts [see Fig. \ref{fig:HomoclinicCollide}(d)] as it approaches the pair of homoclinic orbits associated with the superradiant pole-flip transition F, shown in Fig. \ref{fig:HomoclinicCollide}(e); compare with the image on the Bloch sphere in Fig. \ref{fig:PoleFlip}(b). At this transition the merged chaotic attractor disappears. When viewed for decreasing $\lambda_{+}$, we conclude that this flip transition generates a large symmetric chaotic attractor and the associated chaotic temporal trace of the photon number.

	\subsection{Overall Phase diagram}
	
	\begin{figure}[h]
		\centering
		\includegraphics[width=8.6cm]{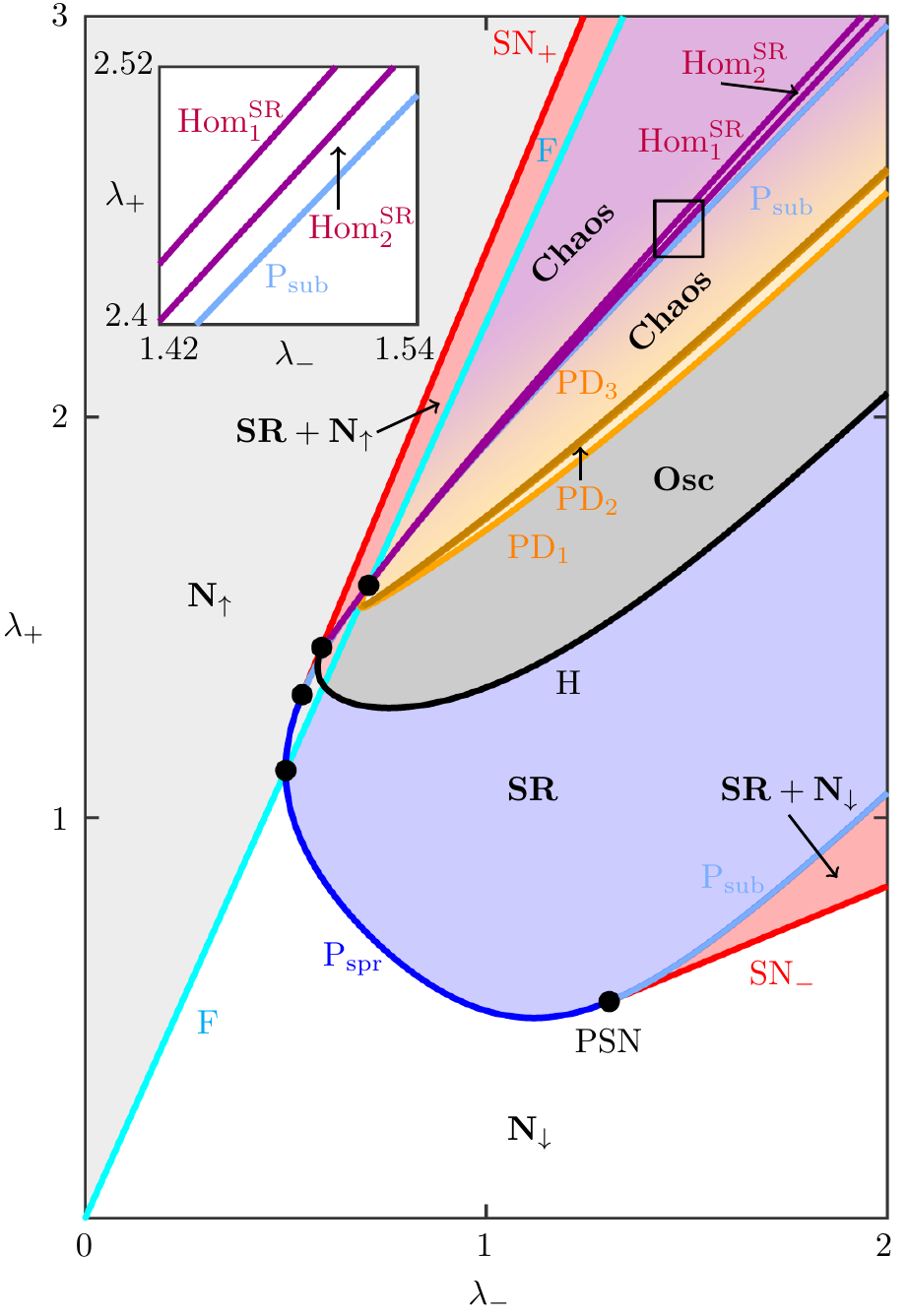}
		\caption{\label{fig:PhaseDiagram}Phase diagram in the $(\lambda_{-},\lambda_{+})$-plane of the unbalanced, open Dicke model for $\kappa=\omega=\omega_0=1$, showing the bifurcation curves from Fig. \ref{fig:Hopf2D} and a selection of bifurcation curves associated with chaotic behavior. Color indicates regions of different behavior: the two normal phase regions, spin-down (white) and spin-up (light gray), are denoted $\mathbf{N}_{\downarrow}$ and $\mathbf{N}_{\uparrow}$, respectively. The superradiant region (blue) is denoted $\mathbf{SR}$, the multistable regions (red) are labelled $\mathbf{SR}+\mathbf{N}_{\downarrow}$ and $\mathbf{SR}+\mathbf{N}_{\uparrow}$, the oscillatory region (gray) is denoted $\mathbf{Osc}$, and the chaotic region (gradient orange to purple) is denoted $\mathbf{Chaos}$, and its shading indicates the transition from isolated chaotic attractors (orange) near the period-doubling curves $\mathrm{PD}_{i}$ to the merged chaotic attractor (purple) near the pole-flip transition curve F.}
	\end{figure}
	
	A two-parameter bifurcation diagram in the $(\lambda_{-},\lambda_{+})$-plane outlining all of the dynamics discussed thus far is presented in Fig. \ref{fig:PhaseDiagram}. Specifically, the phase diagram shown in Fig. \ref{fig:PhaseDiagram} shows the locations of saddle-node bifurcations $\color{black}{\mathrm{SN}_\pm}$, super- and subcritical pitchfork bifurcations $\color{black}{\mathrm{P}_{\mathrm{spr/sub}}}$, and Hopf bifurcations H; compare with Fig.~\ref{fig:Hopf2D}. It also shows the curves of period doubling bifurcations $\color{black}{\mathrm{PD}_{i}}$, and homoclinic bifurcations to superradiant equilibria $\color{black}{\mathrm{Hom}^{\mathrm{SR}}_{i}}$, which are found with numerical continuation; note that we use the same notation $\mathrm{Hom}_{i}$ to refer to the homoclinic orbits and their associated bifurcation curves. The different bifurcation curves divide the $(\lambda_{-},\lambda_{+})$-parameter plane into several distinct regions of different dynamics. These regions are the spin-up or spin-down normal phases $\mathbf{N}_{\uparrow/\downarrow}$ where the photon number tends to zero, the superradiant phase $\mathbf{SR}$ where the atoms collectively emit into the cavity mode, multistable regions $\mathbf{N}_{\uparrow/\downarrow}+\mathbf{SR}$ where there are simultaneous normal and superradiant phases, and the oscillatory phase $\mathbf{Osc}$ where the photon number features simple persistent oscillations. Moreover, there is a region $\mathbf{Chaos}$ which is more complicated. It is bounded by the accumulation of the curves $\mathrm{PD}_{i}$ and the curve F. Inside the region $\mathbf{Chaos}$ we find localized and non-localized chaotic attractors. The transition between the two cases is indicated by graded shading, with the lighter shade indicating two isolated chaotic attractors and darker shade indicating a merged chaotic attractor. As we discussed, the transition between the two cases involves the curves $\mathrm{Hom}_{i}^{\mathrm{S}}$ near the curve $\mathrm{P}_{\mathrm{sub}}$. Figure \ref{fig:PhaseDiagram} also features a number of codimension-two points that act as organizing centers for the creation of codimension-one bifurcation curves; they will be discussed later in the section.
	
	\subsubsection*{Step-through the phase diagram}
	To illustrate typical examples of the dynamics as $\lambda_{-}$ and $\lambda_{+}$ are varied, we follow a relevant closed path through the two-parameter bifurcation diagram Fig. \ref{fig:PhaseDiagram} as we traverse across the parameter plane, beginning in the bottom right-hand region of the spin-down normal phase $\mathbf{N}_{\downarrow}$; see \footnote{See Supplemental Material at \url{https://ksti263.github.io/} for an interactive web version of this phase diagram} for an interactive web version of the bifurcation diagram. 
	
	Consider the path illustrated in Fig. \ref{fig:PhaseDiagramPath}: from $\mathbf{N}_{\downarrow}$ it enters the superradiant region $\mathbf{SR}$ either directly or via the multistable region $\mathbf{SR}+\mathbf{N}_{\downarrow}$. The path then moves counter-clockwise through $\mathbf{Osc}$ and $\mathbf{Chaos}$ to $\mathbf{N}_{\uparrow}$ and back to $\mathbf{N}_{\downarrow}$.
	
	The associated sequence of behavior and bifurcations is as follows:
	
	\begin{figure}[t]
		\centering
		\includegraphics[width=5cm]{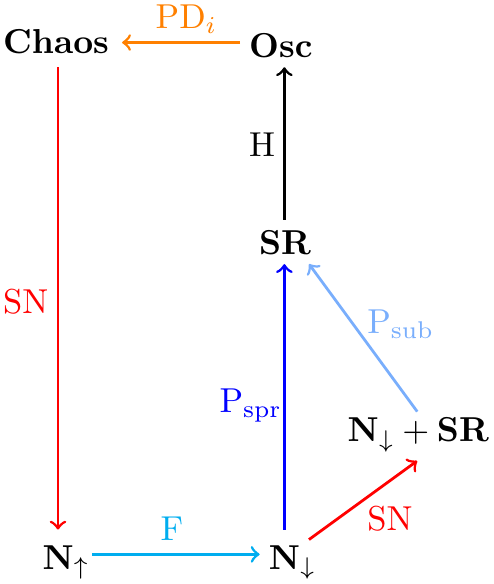}
		\caption{\label{fig:PhaseDiagramPath}Schematic of our step-through the two-parameter bifurcation diagram in Fig. \ref{fig:PhaseDiagram}.}
	\end{figure}
	
	\begin{enumerate}[leftmargin=2.9cm,topsep=0mm,itemsep=-5mm]
		\item[$\textbf{N}_{\downarrow}$]\textit{The system is in the normal phase, with all spins down.} \\
		\item[$\textbf{N}_{\downarrow}\xrightarrow{\mathrm{P}_{\mathrm{spr}}}\textbf{SR}$] \textit{Moving across the supercritical pitchfork bifurcation $\mathrm{P}_{\mathrm{spr}}$ the system undergoes the single-stage transition into superradiance.} \\
		\item[$\textbf{N}_{\downarrow}\xrightarrow{\mathrm{SN}_{-}}\textbf{SR}+\textbf{N}_{\downarrow}$] \makebox[0.3\textwidth][s]{\textit{Alternatively, moving across the}} \\
		\vspace{-1mm}
		\item[$\xrightarrow{\mathrm{P}_{\mathrm{sub}}}\textbf{SR}$\hspace{9mm}] \textit{saddle-node bifurcation $\mathrm{SN}$, the system becomes multistable, with coexistent spin-down normal and superradiant phases, and then becomes entirely superradiant when the subcritical pitchfork bifurcation $\mathrm{P}_{\mathrm{sub}}$ is crossed.} \\
		\item[$\textbf{SR}\xrightarrow{\mathrm{H}}\textbf{Osc}$] \textit{As the system moves across the Hopf bifurcation curve $\mathrm{H}$, the superradiant equilibria bifurcate into stable periodic orbits, and the system enters the oscillatory phase; see Fig. \ref{fig:HopfBifExample}.} \\
		\item[$\textbf{Osc}\xrightarrow{\mathrm{PD}_{i}}\textbf{Chaos}$] \textit{The system undergoes a period-doubling cascade as it moves across the period-doubling curves $\mathrm{PD}_{i}$, forming a pair of asymmetric, localized chaotic attractors; see Fig. \ref{fig:PeriodDoubling}.} \\
		\item[$\textbf{Chaos}$] \textit{Inside of this region the pair of chaotic attractors collide via a sequence of homoclinic bifurcations to the origin $\mathrm{Hom}_{i}^{\mathrm{S}}$ to form a merged chaotic attractor, indicated qualitatively in Fig. \ref{fig:PhaseDiagram} by shading.}\\
		\item[$\textbf{Chaos}\xrightarrow{\mathrm{F}}\textbf{SR}+\textbf{N}_{\uparrow}$] \textit{Transitioning across the curve $\mathrm{F}$, the system undergoes the superradiant pole-flip transition, where the chaotic attractor disappears by contracting onto the pair of homoclinic orbits at $\mathrm{F}$; compare with Fig. \ref{fig:HomoclinicCollide}(d-e). As a result, the system is now multistable and simultaneously in the spin-up normal and superradiant phases.} \\
		\item[$\textbf{SR}+\textbf{N}_{\uparrow}\xrightarrow{\mathrm{SN}_{+}}\textbf{N}_{\uparrow}$]\textit{At the saddle-node bifurcation curve $\mathrm{SN}_{+}$, the superradiant equilibria disappear and the system remains in the spin-up normal phase.} \\
		\item[$\textbf{N}_{\uparrow}\xrightarrow{\mathrm{F}}\textbf{N}_{\downarrow}$] \textit{Moving again across the pole-flip transition curve F, the normal phase transitions from spin-up to spin-down; see Fig. \ref{fig:PoleFlip}(a).}
	\end{enumerate}
	
\begin{figure}[t]
		\centering
		\includegraphics[width=8.6cm]{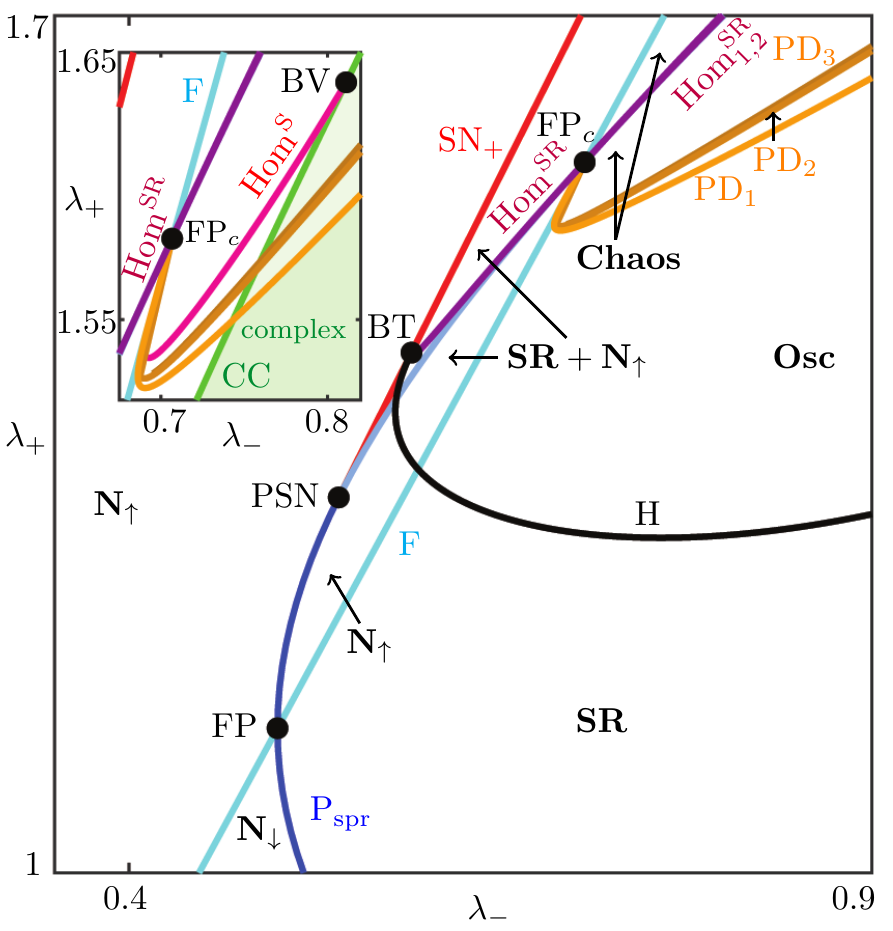}
		\caption{\label{fig:PhaseDiagramZoom}An enlargement of Fig. \ref{fig:PhaseDiagram} accentuating the codimension-two bifurcations. The inset shows the curve (CC) that delineates regions where the equilibrium $\mathrm{N}_{1}^{\mathrm{S}}$ has real or complex conjugate (green shaded region) eigenvalues. Also shown in the inset is a homoclinic bifurcation curve to the South pole equilibrium point ($\mathrm{Hom}^{\mathrm{S}}$).}
	\end{figure}

	A number of codimension-two points are part of the phase diagram in Fig. \ref{fig:PhaseDiagram}. Figure \ref{fig:PhaseDiagramZoom} is an enlargement that shows these near the intersection of the curves $\mathrm{P}_{\mathrm{sub}}$ and F, with a further enlargement as an inset. Codimension-two points act as organizing centers for the codimension-one bifurcation curves of the phase diagram. Apart from the points PSN, FP, and BT already seen in Fig. \ref{fig:Hopf2D}, we now also show in Fig. \ref{fig:PhaseDiagramZoom} the points $\mathrm{FP}_{c}$ and BV, as well as the associated bifurcation curves $\mathrm{P}_{\mathrm{sub}}$, F, and the homoclinic bifurcation curves $\mathrm{Hom}_{i}^{\mathrm{SR}}$, whose associated orbits are homoclinic to $\mathrm{SR}_{1}$, and $\mathrm{Hom}^\mathrm{S}$, whose orbits are homoclinic to $\mathrm{N}_{1}^{\mathrm{S}}$. The point $\mathrm{FP}_c$ in Fig. \ref{fig:PhaseDiagram} and Fig. \ref{fig:PhaseDiagramZoom} is the main organizing center of global bifurcations and chaotic dynamics. Here, the (subcritical) pitchfork bifurcation curve $\mathrm{P}_{\mathrm{sub}}$ crosses the pole-flip transition curve F. This point is responsible for the creation of the period-doubling bifurcation curves $\mathrm{PD}_{i}$ and of infinitely many homoclinic bifurcations to both superradiant and normal phase equilibria. Out of the Bogdanov-Takens point BT emerges a single (superradiant) homoclinic bifurcation curve $\mathrm{Hom}^{\mathrm{SR}}$ which reaches the organizing center $\mathrm{FP}_c$. Coming out of the point $\mathrm{FP}_c$ we expect an infinite number of $m$-homoclinic bifurcations to superradiant equilibria $\mathrm{Hom}_{m}^{\mathrm{SR}}$ and to the South pole of the Bloch sphere, one of which (the same homoclinic orbit as shown in Fig. \ref{fig:HomoclinicCompare}(b)) is shown in the inset of Fig. \ref{fig:PhaseDiagramZoom} as $\mathrm{Hom}^{\mathrm{S}}$. 

	The inset of Fig. \ref{fig:PhaseDiagramZoom} also shows the curve labelled CC where the South pole equilibrium $\mathrm{N}_{1}^{\mathrm{S}}$ transitions from having real to having complex conjugate eigenvalues. At the intersection point BV of the curves $\mathrm{Hom}^{\mathrm{S}}$ and CC, the homoclinic bifurcation transitions from a homoclinic bifurcation to a real saddle to a Shilnikov bifurcation (to a saddle focus with associated decaying oscillations). More specifically, the point BV is a \emph{Belyakov transition point} (see \cite{homburg_homoclinic_2010} for more information), with the eigenvalue condition
	\begin{equation}
	\frac{-v_{s}}{v_{u}}<1,
	\end{equation}
	where $v_{s}=-0.5$ and $v_{u}\approx 1.0985$ are the stable and unstable leading eigenvalues of the South pole equilibrium point. Theory \cite{homburg_homoclinic_2010} tells us that there are infinitely many Shilnikov bifurcation curves to the South pole equilibrium emanating from the Belyakov point BV. We expect that this transition is an accumulation of $m$-homoclinic orbits $\mathrm{Hom}_{m}^{\mathrm{S}}$ emerging from the organizing center $\mathrm{FP}_c$. One such curve, for the homoclinic orbits shown in Fig. \ref{fig:HomoclinicCollide}(b), is shown in the inset of Fig. \ref{fig:PhaseDiagramZoom} (the Shilnikov bifurcation curves emerging on the right-hand side of the CC curve are not shown). This transition should occur for each homoclinic bifurcation curve emerging from the organizing center $\mathrm{FP}_c$, so we expect infinitely many Belyakov points BV: one on each homoclinic curve as it crosses CC.

	\subsection{Effect of detuning}
	
	So far we have considered the case $\omega_0=\omega$ of zero detuning between the effective cavity frequency $\omega$ and the effective atomic frequency $\omega_0$. As we show now, this is not as restrictive as it initially may seem. Namely, the bifurcations we have studied here are \emph{generic} in the sense that, for sufficiently small changes of the parameters $\kappa$, $\omega$, and $\omega_0$, the phase diagram in Fig. \ref{fig:PhaseDiagram} remains qualitatively the same. We now address the natural question of how much of a detuning between $\omega$ and $\omega_0$ is still allowed before the phase diagram changes qualitatively. Note that in the experiments of \cite{zhiqiang_nonequilibrium_2017} $\omega$ and $\omega_0$ depend on the sum and difference, respectively, of the frequencies of the driving lasers, so either can be changed quite easily in an experiment. To be definite, we consider changing the effective atomic frequency $\omega_0$, while holding the effective cavity frequency $\omega$ constant. 

We find that the relative positions of organizing centers (codimension-two points) of the phase diagram in Fig. \ref{fig:PhaseDiagram} do not change qualitatively when $\omega_0$ is decreased over the entire interval $0<\omega_0\leq\omega$. More specifically, it follows from the expression \eqref{eqn:PoleFlip} for its slope that, as $\omega_0$ approaches zero, the pole-flip transition curve F approaches the diagonal in the $(\lambda_{-},\lambda_{+})$-plane where $\lambda_{-} = \lambda_{+}$. Similarly, according to expression \eqref{eqn:Pitchfork}, the curve of pitchfork bifurcations approaches this diagonal as well as $\omega_0 \to 0$. In other words, the curves F and P collapse onto the diagonal in this limit. Nevertheless, we find that F and P do not change their relative positions until the limit $\omega_0 = 0$ is reached, and the same is true for the curves emerging from their intersection point, the organizing center $\mathrm{FP}_c$.

When $\omega_0$ is increased from $\omega_0=\omega$ the relative positions of the organizing centers of phase diagram in Fig. \ref{fig:PhaseDiagram} do not change qualitatively over the interval $\omega<\omega_0\leq\omega_0^*$, where $\omega_0^*=\sqrt{\omega^2+\kappa^2}$. At this critical frequency $\omega_0^*$ there is a qualitative change of the phase diagram because then the expression \eqref{eqn:PoleFlip} for the slope of pole-flip transition curve F becomes
	\begin{equation}
	\frac{\omega+\sqrt{\omega^2+\kappa^2}}{\kappa},
	\end{equation}
	which, according to  \eqref{eqn:SaddleNode}, is exactly the ($\omega_0$-independent) slope of the saddle-node curve $\mathrm{SN}_{+}$. Thus, at the critical frequency $\omega_0^*$ the curve $\mathrm{SN}_{+}$  coincides with F, which corresponds to the codimension-three situation where the points FP, $\mathrm{FP}_{c}$, PSN, and BT all coincide as a unique point of tangency between the pitchfork bifurcation curve $\mathrm{P}_{\mathrm{sub}}$ and the pole-flip transition curve F. For $\omega_0 > \omega_0^*$ the relative position of the points FP and $\mathrm{FP}_c$ along the curve $\mathrm{P}_{\mathrm{sub}}$ has changed. This means that these two codimension-two points effectively move through each other at $\omega_0 = \omega_0^*$, leading to a qualitative change of the phase diagram. A detailed analysis of the phase diagram in the ($\lambda_{-},\lambda_{+}$)-plane for $\omega_0 > \omega_0^*$ is beyond the scope of this work, but we conjecture that it does not change qualitatively until the limit $\omega_0 \to \infty$ is reached, when the pole-flip transition curve F also reaches the diagonal where $\lambda_{-} = \lambda_{+}$. 

We conclude that the phase diagram as shown in Fig. \ref{fig:PhaseDiagram} is representative of the dynamics if $0<\omega_0<\omega_0^*$, which constitutes a considerable range of detunings. Because of the parameter exchange symmetry (\ref{eqn:ParameterSymmetry}), the same is true when $-\omega_0^*<\omega_0<0$, subject to an exchange of the coupling strengths $\lambda_{-}\leftrightarrow\lambda_{+}$; this also implies a reversal of the roles of the North and South pole of the Bloch sphere, i.e., the pitchfork bifurcation to superradiance takes place at the North pole in this case. 

We finally remark that the critical frequency $\omega_0^*$ is characterized by a specific relationship between the cavity decay rate $\kappa$ and the effective atomic and cavity frequencies $\omega_0$ and $\omega$. That is, if one sees the light field amplitude $\alpha$ and the atomic polarization $\beta$ as isolated oscillators (i.e., zero coupling) then $\omega_0^*$ separates the situation where the frequency of the atoms is faster or slower, respectively, than the combined decay and frequency of the light field. It is an interesting observation that this relation has consequences for the overall organization of the phase diagram of the unbalanced, open Dicke model, namely by generating a codimension-three bifurcation of the nonlinear dynamics.

	
	\section{Conclusion and Outlook}\label{Sect:Conclusion}
	
	We have analyzed phase transitions in the unbalanced, open Dicke model in the semiclassical regime with a dynamical systems approach. In doing so, we examined how unbalanced coupling leads to a splitting of the emergence of superradiance and the destruction of the normal phase in the semiclassical description, giving rise to the phase coexistence studied in \cite{soriente_dissipation-induced_2018}. We have also investigated the emergence of superradiant oscillations in terms of Hopf bifurcations that the system undergoes.
	
	During the analysis, we discovered a transition between the two types of normal phase configurations, spin-up and spin-down, which we have called the pole-flip transition. We found that at the point of transition, an infinite number of energy-conserving periodic orbits foliate the Bloch sphere. This can occur in two ways: the ``normal" pole-flip transition where the periodic orbits oscillate about the poles, or the ``superradiant" case where periodic orbits develop also about superradiant equilibria. Explicit expressions for the pole-flip transition and the conserved energy were determined. 
	
	We have also demonstrated the onset of chaos in the system arising from period-doubling cascades of periodic orbits created at the Hopf bifurcations. These create two symmetry-related localized chaotic attractors, which can then collide via a sequence of homoclinic bifurcations to form a single symmetric and non-localized chaotic attractor as parameters are varied. We also showed that this type of symmetric chaotic attractor can arise when the foliation of periodic orbits that arise from the superradiant pole-flip transition is perturbed, which constitutes a completely different mechanism for the creation of chaotic behavior. 

The different regions of possible behaviours and transition curves between them have been represented as a phase diagram in the $(\lambda_{-},\lambda_{+})$-plane, which clearly identifies where superradiance, multistability, oscillations, and chaos can be found. We have shown how the overall phase diagram is organized by a number of codimension-two bifurcation points, from where different transition curves meet and/or emerge. In fact, the point $\mathrm{FP}_c$, where the curves of symmetry breaking and the pole-flip transition cross, acts as the main organizing center that generates the chaotic dynamics that can be observed in the system. We showed that the relative positions of the different codimension-two points do not change when the effective cavity and atomic frequencies are detuned over a very large range, up to a critical value that we determined explicitly. This implies that the phase diagram presented here is valid well beyond the case of zero detuning for which it was computed. 

The generalization of unbalanced coupling, which may seem like a simple generalization, reveals such rich and complicated structure of additional dynamics. It is quite surprising given that considering unbalanced coupling does not add any new, explicitly nonlinear terms to the Dicke Hamiltonian; it simply splits the contribution of the interaction term of the standard Dicke model to two independently varying terms. Indeed, the dynamics of the standard Dicke model can be found along the diagonal of the $(\lambda_{-},\lambda_{+})$-plane, and all the additional behaviours we found require a quantifiable amount of unbalancing.

There are clearly some promising directions for further investigation of the semiclassical approximation of the unbalanced, open Dicke model. First of all, ongoing work is devoted to a more detailed study of the bifurcation sequence involved in the collision of two localized chaotic attractors to form a non-localized chaotic attractor. Another task is to determine what the phase diagram looks like beyond the critical value of the detuning between the effective cavity and atomic frequencies. We have already presented some ingredients in terms of the relative positions of organizing centers, but a proper answer will require a substantial bifurcation analysis in the spirit of the one presented here. Finally, it is an interesting mathematical challenge to study the exact nature of the main organizing center $\mathrm{FP}_c$, which involves the existence of a two-dimensional surface that connects both poles of the Bloch sphere, and which is foliated by periodic orbits and a pair of homoclinic orbits. Since such a foliation is rather special, in the sense that there is a conserved quantity, an important related question is how this situation, which occurs along an entire line in the $(\lambda_{-},\lambda_{+})$-plane, can be unfolded. This will require the consideration of additional decay channels, such as (either individual or collective) dissipation of the spins as considered in \cite{Gelhausen_manybody_2017,shchadilova_fermionic_2020}. In the semiclassical regime, the addition of atomic decay means that angular momentum is no longer conserved so the atomic dynamics are freed from the Bloch sphere, adding an additional degree of freedom to the dynamics.

Our study has also shown that regions in the $(\lambda_{-},\lambda_{+})$-plane corresponding to the different behaviors described in this work are sufficiently large that it should be possible to reach them experimentally. This will obviously require the implementation of sufficiently unbalanced coupling strengths, among other experimental challenges. The possibility of future experiments naturally raises many questions regarding the correspondence of the semiclassical results presented here with the fully quantum mechanical model for finite $N$. Specifically, the region of multistability, where there is coexistence of the normal and superradiant phases, could provide an avenue for studying the correspondence of quantum to semiclassical dynamics since basins of different attractors are typically not well defined in a quantum mechanical context. In addition, any oscillations in the quantum model will typically be washed out of the master equation description by its statistical nature. A possible approach to analyzing oscillations and chaos in the fully quantum mechanical model is to utilize a quantum trajectory method \cite{breuer_theory_2007,daley_quantum_2014,carmichael_statistical_2007}. In particular, our demonstration that its semiclassical approximation of the unbalanced, open Dicke model possesses rich chaotic dynamics implies that studying its quantum mechanical counterpart could provide an interesting context in which to study the onset of quantum chaos.

	
	\section{Acknowledgements}
	The authors would like to thank Stuart Masson for many insightful and interesting discussions.
	
	
	\appendix*
	\section{Superradiant bifurcation curves with parameter change}
	
	Equations (\ref{eqn:Pitchfork}) and (\ref{eqn:Dimer}) are invariant under the parameter exchange $\lambda_{+}\leftrightarrow\lambda_{-}$, which produces the reflection symmetry of the superradiant phase diagram Fig. \ref{fig:SprRad2D} about the diagonal. This suggests a slightly more `natural' parameter choice to express the coupling strengths as
	\begin{equation}
	\mu_{\pm} = \lambda_{+}\pm\lambda_{-}.
	\end{equation}
	Expressed in these parameters, the locations of the pitchfork bifurcations satisfy
	\begin{equation}
	\mu_{+}^2\mu_{-}^2 - \omega\omega_0(\mu_{+}^2 + \mu_{-}^2) + \omega_0^2(\kappa^2 + \omega^2) = 0.
	\end{equation}
	This equation is quadratic in $\mu_{+}$, the solution is
	\begin{equation}
	\mu_{+} = \pm\sqrt{\frac{\omega\omega_0\mu_{-}^2-\omega_0^2(\kappa^2 + \omega^2)}{\mu_{-}^2-\omega\omega_0}}.
	\end{equation}
	The implicit equation that determines the locations of the saddle-node bifurcations is
	\begin{equation}
	-\frac{\omega^2}{4}\left( \mu_{+}^2 + \mu_{-}^2 \right) + \left( \kappa^2 + \frac{\omega^2}{2} \right)\mu_{+}^2\mu_{-}^2 = 0,
	\end{equation}
	which is quadratic in $\mu_{+}^2$, and has the solution
	\begin{equation}
	\mu_{+} = \pm\left(\frac{\sqrt{2\kappa^2 + \omega^2 - 2\kappa\sqrt{\kappa^2 + \omega^2}}}{\omega}\right)\mu_{-}.
	\end{equation}
	
	
	\bibliography{SGKP_Dicke}
	
\end{document}